\documentclass[useAMS,usenatbib,usegraphicx]{mn2e}

\usepackage[scriptsize]{subfigure}
\usepackage{multirow}
\usepackage{enumerate}
\usepackage{amssymb,amsmath,bm}%split
\title[Estimating stellar atmospheric parameters based on LASSO and support-vector regression]{Estimating stellar atmospheric parameters based on LASSO and support-vector regression}
\author[Yu Lu and Xiangru Li]{Yu Lu,Xiangru Li\thanks{Email:
xiangru.li@gmail.com(X. Li)}\\
School of Mathematical Sciences, South China Normal University, Guangzhou 510631, China; %xiangru.li@gmail.com
}

\begin{document}
%\begin{CJK}{GBK}{song}%ok
%\date{Accepted ???? December 15. Received 2014 December 14; in original form 1988 October 11}

\pagerange{\pageref{firstpage}--\pageref{lastpage}} \pubyear{2015}

\maketitle

\label{firstpage}

\begin{abstract}
A scheme for estimating atmospheric parameters T$_\texttt{eff}$, log$~g$, and [Fe/H] is proposed on the basis of Least Absolute Shrinkage and Selection Operator (LASSO) algorithm and Haar wavelet. The proposed scheme consists of three processes. A spectrum is decomposed using the Haar wavelet transform and low-frequency components at the fourth level are considered as candidate features. Then, spectral features from the candidate features are detected using the LASSO algorithm to estimate the atmospheric parameters. Finally, atmospheric parameters are estimated from the extracted spectral features using the support-vector regression (SVR) method. The proposed scheme was evaluated using three sets of stellar spectra respectively from Sloan Digital Sky Survey (SDSS), Large Sky Area Multi-object Fiber Spectroscopic Telescope (LAMOST), and Kurucz's model, respectively. The mean absolute errors are as follows: for 40~000 SDSS spectra, 0.0062 dex for log~T$_\texttt{eff}$ (85.83 K for T$_\texttt{eff}$), 0.2035 dex for log$~g$ and 0.1512 dex for [Fe/H]; for 23963 LAMOST spectra, 0.0074 dex for log~T$_\texttt{eff}$ (95.37 K for T$_\texttt{eff}$), 0.1528 dex for log~$g$, and 0.1146 dex for [Fe/H]; and for 10469 synthetic spectra, 0.0010 dex for log T$_\texttt{eff}$(14.42K for T$_\texttt{eff}$), 0.0123 dex for log~$g$, and 0.0125 dex for [Fe/H].
\end{abstract}


\section{Introduction}\label{Sec:introduction}

The implementation of large-scale and deep sky survey programs, such as the Sloan Digital Sky Survey \citep[SDSS:][]{Journal:York:2000,Journal:Ahn:2012}, Gaia-ESO Survey  \citep[GES:][]{Journal:Gilmore:2012,Journal:Randich:2013} and the Large Sky Area Multi-object Fiber Spectroscopic Telescope \citep[LAMOST/Guoshoujing Telescope;][]{Journal:Zhao:2006,Journal:Cui:2012}, has resulted in the collection of a large amount of stellar spectra. The bulk of this stellar spectral information necessitates utilisation of a fully automated characterisation process.

In particular, estimating the effective temperature, surface gravity, and metallicity from stellar spectra remains a fundamental problem \citep{Wu2011,Song2012}. For example, \citet{Journal:Muirhead:2012} investigated the estimation of the effective temperature $T_\texttt{eff}$ and metallicity [M/H] for late-K and M-type planet-candidate host stars from the K-band spectra released by the Kepler Mission. In this study, a surface fitting method was used along with three spectral indices, namely the equivalent widths of NaI (2.210 $\mu$m) and CaI (2.260 $\mu$m) lines and an index describing the flux change between three 0.02$\mu$m wide bands centred at 2.245, 2.370, and 2.080 $\mu$m. \citet{Journal:Koleva:2009} developed a full-spectrum fitting package, ULySS (University of Lyon Spectroscopic analysis Software), and explored its application in spectral parameterization. \citet{Journal:Manteiga:2010} parameterised stellar spectra by extracting features based on Fourier analysis and wavelet decomposition as well as by constructing a mapping from feature space to parameter space using a forward neural networks (FNN) with three layers.

An automated estimation system for atmospheric parameters from a stellar spectrum is also referred to as a spectrum-parameterisation system (SPS) in related studies. An SPS system consists of two key processes, namely feature extraction and parameter estimation. Feature extraction determines representation of spectral information. Principal component analysis \citep[PCA:][]{Journal:Fiorentin:2007,Journal:Bu:2015} and fast Fourier transform \citep[FFT:][]{Journal:Manteiga:2010} are the two most frequently used feature extraction methods. Parameter estimation constructs a mapping from the extracted features of a spectrum to its atmospheric parameters. Some commonly used estimation methods are FNN \citep{Journal:Fiorentin:2007}, Gaussian process \citep[GP:][]{Journal:Bu:2015}, support-vector regression \citep[SVR:][]{Journal:Li:2014}, etc.

This work describes a scheme for estimating atmospheric parameters from the stellar spectrum. The three processes of the proposed scheme are performed as follows. Several candidate features are initially obtained by transforming a spectrum into a low-frequency space using the Haar wavelet transform. Then, some representative candidate features are selected as spectral features using the Least Absolute Shrinkage and Selection Operator (LASSO) algorithm \citep{Journal:Tibshirani:1996}. Finally, atmospheric parameters are estimated from extracted features using a regression method.

High-frequency components are removed in the first process. These components are usually more affected by noise than low-frequency components, which will be discussed further in section \ref{Sec:Exper_Disc:Filtering_Selection}. The second process reduces the number of spectral features efficiently, a feature closely related to the estimation efficiency and exploration of representative features (Section \ref{Sec:Exper_Disc:Filtering_Selection}). In the third process, the spectrum parameterization problem is investigated using four typical estimation methods: the radial basis function neural network \citep[RBFNN:][]{Journal:Schwenker:2001}, SVR \citep{Tec:Chang:2001,Journal:Schokopf:2002,Journal:Smola:2004}, $K$-nearest neighbor regression \citep[KNNR:][]{Journal:Altman:1992} and least square regression \citep[LSR:][]{Book:James:2013}. Experimental results indicate that SVR is superior to the other three methods.

This article is organised as follows. Section \ref{Sec:Data} describes some applied experimental data. After introducing four regression methods in Section \ref{Sec:Estimator_Evaluation}, Section \ref{Sec:Feature_Extraction} describes a scheme for extracting spectral features. In Section \ref{Sec:Exper_Disc}, the proposed scheme is evaluated experimentally. Finally, Section \ref{Sec:Conclusion} concludes this work

\section{Data}\label{Sec:Data}
The proposed scheme was evaluated by performing experiments using three data sets: 33~963 actual spectra from LAMOST, 50~000 actual spectra from SDSS and 18~969 synthetic spectra computed from Kurucz¡¯s model. Actual spectra usually carry some disturbances from noise and pre-processing imperfections.

The proposed scheme is a statistical learning method. It basically involves the discovery of mapping from stellar spectra to their atmospheric parameters using empirical data, which is referred to as a training set in machine learning. At the same time, the performance of the mapping discovered should be evaluated objectively. Therefore, an independent set of stellar spectra is needed for this evaluation. This independent set is usually referred to as a test set. Thus, in each experiment, stellar spectra are split into two subsets, namely the training and test sets.

\subsection{Actual spectra from SDSS}\label{Sec:Data:SDSS}

A set of 50~000 stellar spectra and their physical parameters collected from SDSS are selected \citep{Abazajian2009,Yanny2009}. The selected spectra span the ranges of [4088, 9740] K for T$_\texttt{eff}$, [1.015, 4.998] dex for surface gravity and [-3.497, 0.268] dex for [Fe/H]. All stellar spectra are shifted to their rest frames (zero radial velocity) using the radial velocity provided by the SDSS/the Sloan Extension for Galactic Understanding and Exploration (SEGUE) Spectroscopic Parameter Pipeline \citep[SSPP:][]{Journal:Beers:2006,Journal:Lee:2008:a,Journal:Lee:2008:b,Journal:Lee:2011,Journal:Prieto:2008,Journal:Smolinski:2011} and rebinned to a maximal common log(wavelength) range [3.581862, 3.963961] with a sampling step of 0.0001. The common wavelength range is approximately [3818.23, 9203.67]\AA. The SDSS training and test sets are labeled as $S_{tr}^{SD}$ and $S_{te}^{SD}$, respectively. Without special instructions, the sizes of $S_{tr}^{SD}$ and $S_{te}^{SD}$ are 10~000 and 40~000, respectively.

\subsection{Actual spectra from LAMOST}\label{Sec:Data:LAMOST}

A set of 33~963 LAMOST stellar spectra \citep{Luo2012} and their physical parameters from LAMOST pipeline are selected based on two signal-to-noise ratio (SNR) constraints: namely SNR$_g \geq 20$ and SNR$_r \geq 20$ in $g-$band and $r-$band \citep{Luo2012} \footnote{This constraint was not used on SDSS data.}. This constraint is proposed by considering the current quality stability of the LAMOST spectra. The LAMOST spectra are also shifted to their rest frames using the radial velocity provided by LAMOST SSPP and rebinned to their maximal common log(wavelength) range [3.5845, 3.9567] with a sampling step of 0.0001. The common wavelength range is approximately [3841.49, 9051.07]\AA. All LAMOST spectra span the ranges [3853.2, 9927] K for T$_\texttt{eff}$, [0.8920, 4.9959]dex for log~$g$ and [-2.3280, 0.9360] dex for [Fe/H]. The LAMOST training and test sets are labeled as $S_{tr}^{LA}$ and $S_{te}^{LA}$, respectively. $S_{tr}^{LA}$ consists of 10,000 stellar spectra, while $S_{te}^{LA}$ consists of 23,963 spectra.

\subsection{Synthetic spectra}\label{Sec:Data:KURUCZ}

A set of 18~969 synthetic spectra are calculated from Kurucz's NEWODF (new opacity distribution function) models \citep{Journal:Castelli:2003} using the SPECTRUM (v2.76) package \citep{Con:Gray:1994} as well as 830~828 atomic and molecular lines (contained in two files, luke.lst and luke.nir.lst). These atomic and molecular data are stored in the file stdatom.dat, which includes solar atomic abundances from \citet{Journal:Grevesse:1998}. The SPECTRUM package and the three data files can be downloaded from website \footnote{http://www.appstate.edu/~grayro/spectrum/spectrum.html}.

The grids of the synthetic stellar spectra span parameter ranges [4000,9750] K for $T_\texttt{eff}$ (45 values, stepsizes of 100 K between 4000 K and 7500 K and 250 K between 7750 K and 9750 K), [1, 5] dex for log$~g$ (17 values, stepsize of 0.25 dex) and [-3.6, 0.3] dex for [Fe/H] (27 values, stepsizes of 0.2 dex between -3.6 dex and -1 dex and 0.1 dex between -1 dex and 0.3 dex). The synthetic stellar spectra are also split into two subsets: a training set $S_{tr}^{SY}$ and a test set $S_{te}^{SY}$ consisting of 10~469 and 8~500 spectra, respectively.

\section{Estimation Models and Evaluation Methods}\label{Sec:Estimator_Evaluation}
Here, $\bm{x}$ represents a description of a stellar spectrum, while $y$ is the atmospheric parameter $T_\texttt{eff}$, log~$g$, or [Fe/H] of $\bm{x}$.
\subsection{Estimation models}\label{Sec:Estimator_Evaluation:Estimators}
The spectrum parameterization problem is to recover the mapping $f$:
\begin{equation}\label{Equ:parameterization_problem}
y = f(\bm{x})
\end{equation}
from a set of empirical data (training set).

In this work, $f$ is estimated using four typical regression methods: RBFNN \citep{Journal:Schwenker:2001}, SVR \citep{Tec:Chang:2001,Journal:Schokopf:2002,Journal:Smola:2004}, KNNR \citep{Journal:Altman:1992} and LSR \citep{Book:James:2013}. RBFNN and KNNR are non-linear regression methods, while LSR is a linear regression method. SVR can be implemented with a Gaussian kernel and a linear kernel. These two implementation cases are labelled as SVR$_G$ and SVR$_l$, respectively. Therefore, SVR$_G$ is a nonlinear method whereas SVR$_l$ is a linear method.

\subsection{Evaluation criteria}\label{Sec:Estimator_Evaluation:Evaluations}
Suppose that $\hat{f}$ is an estimate of the spectrum parameterization mapping $f$ and $S = \{(\bm{x},y)\}$ is a data set, where $\bm{x}$ is a representation of a stellar spectrum and $y$ is an atmospheric parameter of the corresponding star. The data set $S$ can be a training or test set.

For convenient comparison with related reports, this study evaluates the performance of an estimate $\hat{f}$ using three methods: mean error (ME), mean of absolute error (MAE) and standard deviation (SD). These three evaluation methods are widely used in related studies \citep{Journal:Fiorentin:2007,Journal:Jofre:2010,Journal:Tan:2013}.

Using the data set $S$, the three evaluation criteria can be defined as follows:
\begin{equation}\label{Equ:ME}
  \texttt{ME}=\frac{1}{n}\sum_{(\bm{x}, y) \in S}(y-\hat{f}(\bm{x})),
\end{equation}
\begin{equation}\label{Equ:MAE}
  \texttt{MAE}=\frac{1}{n}\sum_{(\bm{x}, y) \in S}|y-\hat{f}(\bm{x})|,
\end{equation}
\begin{equation}\label{Equ:SD}
  \texttt{SD}=\sqrt{\frac{1}{n}\sum_{(\bm{x}, y) \in S}(y-\hat{f}(\bm{x}))^{2}},
\end{equation}
where $n$ is the amount of elements in $S$.

\section{FEATURE EXTRACTION}\label{Sec:Feature_Extraction}
\subsection{Extract Candidate Features}\label{Sec:Feature_Extraction:Haar}
Atmospheric parameters $T_\texttt{eff}$, log$~g$, and [Fe/H] show an evident non-linear dependence on stellar spectra \citep[Table 6, Table 10, Table 11 in ][]{Journal:Li:2014}. Therefore, the spectrum parameterization problem is investigated by non-linearly transforming a spectrum before estimating atmospheric parameters. In this work, stellar spectra are transformed using a Haar wavelet \citep{Book:Mallat:2009} and decomposed into a series of components with different wavelengths and frequencies (time-frequency localization).

High-frequency components are usually more affected by noise than low-frequency components. Thus, this work obtains candidate features by removing high-frequency components. This process will be discussed further in Section \ref{Sec:Exper_Disc:Filtering_Selection}.

In addition to the Haar wavelet, multiple alternatives for a non-linear transformation \citep{Book:Mallat:2009,Book:Daubechies:1992}, such as the Coiflets, Daubechies, Symmlet and biorthogonal wavelets, are available based on wavelet transform.

\subsection{Refining the candidate features}\label{Sec:Feature_Extraction_Extraction:Refining_lasso}
Experiments indicate the presence of many redundancies in the extracted candidate features (Section \ref{Sec:Exper_Disc:Filtering_Selection}). Therefore, this work proposes a scheme for detecting spectral features from the extracted candidate features using the LASSO algorithm \citep{Journal:Tibshirani:1996}.

Let $S_{tr} = \{(\bm{x},y)\}$ be a training set (Section \ref{Sec:Data}), where $\bm{x}= (x_1, \cdots, x_m)$ represents a stellar spectrum based on its candidate features, $y$ is an atmospheric parameter of the corresponding star and $m$ is a positive integer. The LASSO algorithm selects features using the following model:
\begin{equation}\label{Equ:LASSO}
   \hat{\bm{w}} = arg\min\limits_{\bm{w}}{\{\sum_{(\bm{x}, y) \in S}{(y -\sum^{m}_{i=1}w_{i}x_i)^2} + \lambda \sum^{m}_{i=1}\|w_{i}\|}\},
\end{equation}
where $\lambda> 0$ is a preset parameter. In this model, only a few $\hat{w}_i$ will be non-zero, and the variables $x_i$ with a non-zero $\hat{w}_i$ are selected as spectral features. The parameter $\lambda$ controls the amount of non-zero parameters $\hat{w}_i$ or, equivalently, the number of detected features. In this work, the parameter $\lambda$ is estimated by tenfold cross validation \citep{Journal:Tibshirani:1996,Jouranl:Sjostrand:2005}.

Based on the training set from SDSS (Section \ref{Sec:Data:SDSS}), 17 spectral features are detected for T$_\texttt{eff}$, 24 spectral features for log$~g$, and 25 features for [Fe/H] (Table \ref{Tab:features}).

\begin{table*}
\centering
\caption{ Extracted features for estimating atmospheric parameters. WP is the wavelength position represented by a two-dimensional vector [$a$, $b$], where $a$ and $b$ are the starting and ending wavelengths(\AA). }
\begin{tabular}{ c c c c c c c c}
\hline %\hline
\hline
\multicolumn{8}{c}{(a) Extracted features for estimating T$_\texttt{eff}$.} \\
\hline
Label    &          WP({\AA})   &
Label    &          WP({\AA})   &
Label    &          WP({\AA})   &
Label    &          WP({\AA}) \\
\hline
T$_1$    &[3932.439,  3946.045] &
T$_2$    &[4217.567,  4232.159] &
T$_3$    &[4780.375,  4796.915] &
T$_4$    &[4851.343,  4868.128] \\
T$_5$    &[5033.406,  5050.821] &
T$_6$    &[5070.631,  5088.175] &
T$_7$    &[5108.131,  5125.804] &
T$_8$    &[5126.984,  5144.723] \\
T$_9$    &[5145.908,  5163.712] &
T$_{10}$ &[5164.901,  5182.771] &
T$_{11}$ &[5203.098,  5221.100] &
T$_{12}$ &[6562.389,  6585.094] \\
T$_{13}$ &[8524.364,  8553.857] &
T$_{14}$ &[8650.914,  8680.845] &
T$_{15}$ &[8747.058,  8777.321] &
T$_{16}$ &[8844.270,  8874.870] \\
T$_{17}$ &[9008.697,  9039.866] &
& &
& &
&\\
\hline
\multicolumn{8}{c}{(b) Extracted features for estimating log$~g$.} \\
\hline
Label    &          WP({\AA})   &
Label    &          WP({\AA})   &
Label    &          WP({\AA})   &
Label    &          WP({\AA}) \\
\hline
L$_1$    &[3818.229, 3831.440]&
L$_2$    &[3889.215, 3902.672]&
L$_3$    &[3932.439, 3946.045]&
L$_4$    &[4095.076, 4109.245]\\
L$_5$    &[4295.978, 4310.841]&
L$_6$    &[4540.064, 4555.772]&
L$_7$    &[4556.821, 4572.587]&
L$_8$    &[4573.640, 4589.464]\\
L$_9$    &[4658.671, 4674.789]&
L$_{10}$ &[4833.503, 4850.226]&
L$_{11}$ &[4851.343, 4868.128]&
L$_{12}$ &[4869.249, 4886.096]\\
L$_{13}$ &[4887.221, 4904.130]&
L$_{14}$ &[4923.365, 4940.399]&
L$_{15}$ &[5164.901, 5182.771]&
L$_{16}$ &[5183.964, 5201.900]\\
L$_{17}$ &[5222.302, 5240.371]&
L$_{18}$ &[5241.577, 5259.712]&
L$_{19}$ &[5280.341, 5298.611]&
L$_{20}$ &[5299.831, 5318.167]\\
L$_{21}$ &[5319.392, 5337.796]&
L$_{22}$ &[5418.287, 5437.033]&
L$_{23}$ &[5498.725, 5517.750]&
L$_{24}$ &[6562.389, 6585.094]\\
\hline
\multicolumn{8}{c}{(c) Extracted features for estimating [Fe/H].} \\
\hline
Label    &          WP({\AA})   &
Label    &          WP({\AA})   &
Label    &          WP({\AA})   &
Label    &          WP({\AA}) \\
\hline
F$_1$    &[3932.439, 3946.045]&
F$_2$    &[3990.819, 4004.626]&
F$_3$    &[4005.549, 4019.407]&
F$_4$    &[4020.333, 4034.242]\\
F$_5$    &[4035.172, 4049.133]&
F$_6$    &[4506.735, 4522.327]&
F$_7$    &[4607.465, 4623.406]&
F$_8$    &[4745.282, 4761.700]\\
F$_9$    &[4780.375, 4796.915]&
F$_{10}$ &[4798.019, 4814.620]&
F$_{11}$ &[4815.729, 4832.390]&
F$_{12}$ &[4851.343, 4868.128]\\
F$_{13}$ &[4869.249, 4886.096]&
F$_{14}$ &[4941.536, 4958.633]&
F$_{15}$ &[4959.775, 4976.935]&
F$_{16}$ &[5051.984, 5069.463]\\
F$_{17}$ &[5108.131, 5125.804]&
F$_{18}$ &[5241.577, 5259.712]&
F$_{19}$ &[5260.924, 5279.126]&
F$_{20}$ &[5280.341, 5298.611]\\
F$_{21}$ &[5299.831, 5318.167]&
F$_{22}$ &[5398.362, 5417.039]&
F$_{23}$ &[5438.285, 5457.101]&
F$_{24}$ &[8524.364, 8553.857]\\
F$_{25}$ &[8650.914, 8680.845]\\

\hline
\end{tabular}\label{Tab:features}
\end{table*}

\section{Experiments and discussion}\label{Sec:Exper_Disc}

\subsection{Performance for SDSS spectra}\label{Sec:Exper_Disc:SDSS}

From the detected features in Table \ref{Tab:features}, a spectrum parameterization model can be learned from the training set $S_{tr}^{SD}$ (Section \ref{Sec:Data:SDSS}). The performance obtained using the test set $S_{te}^{SD}$ is presented in Table \ref{table:errors:SDSS}.

\begin{figure*}
  \centering
  \subfigure[log T$_\texttt{eff}$  ]
    { \includegraphics[height=1.5in,width=2.2in]{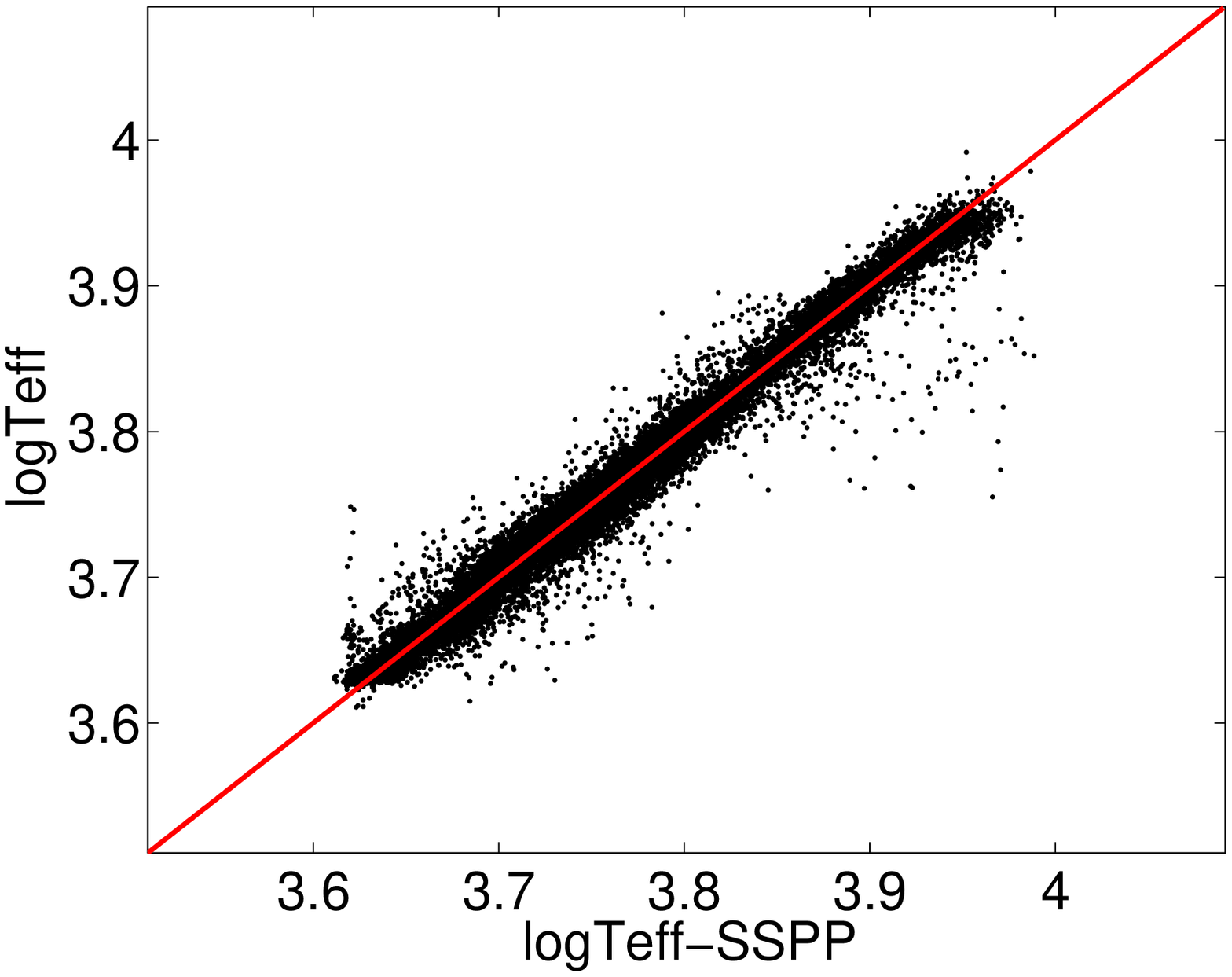} }
  \subfigure[log$~g$]
    { \includegraphics[height=1.5in,width=2.2in]{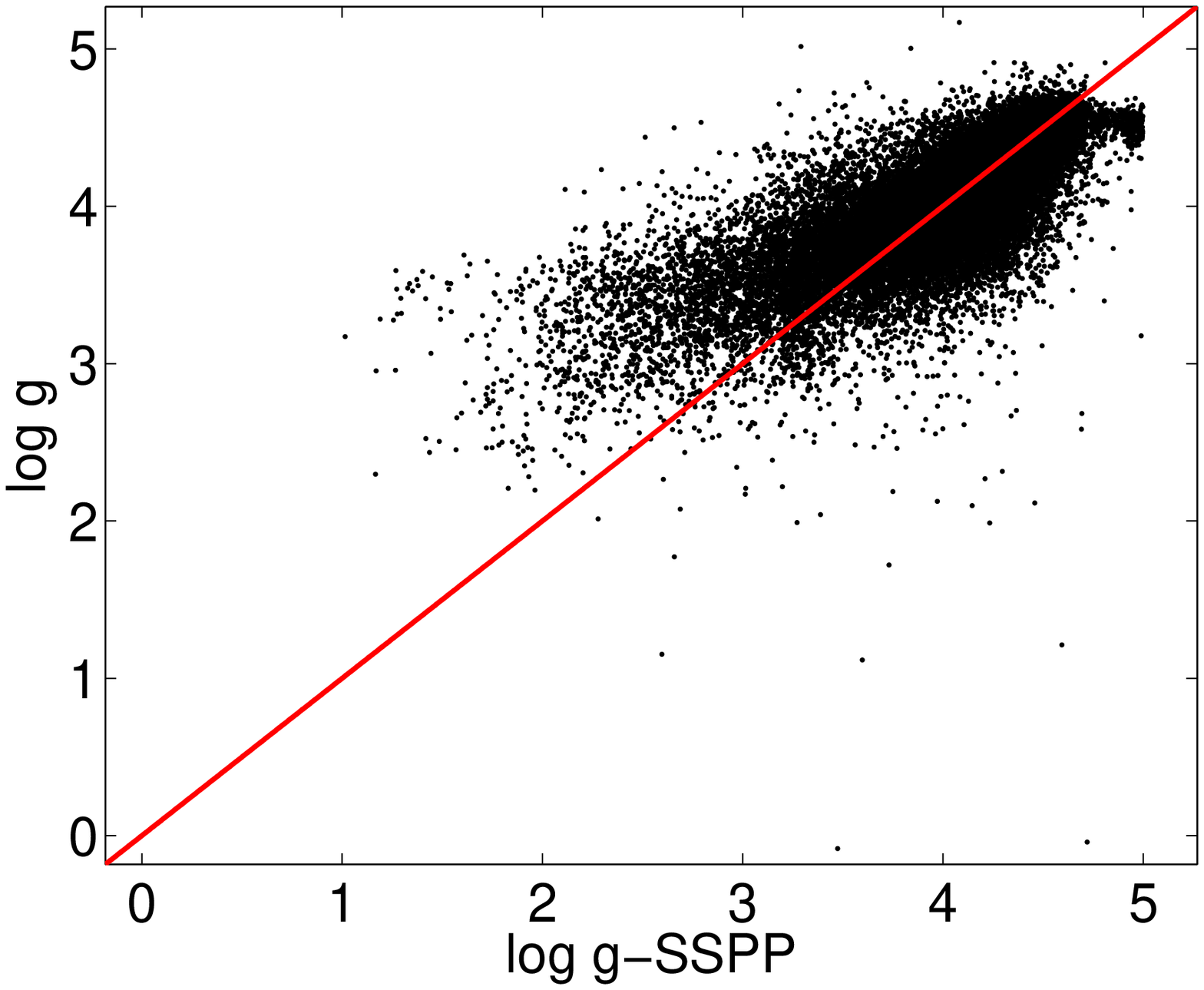} }
  \subfigure[{[Fe/H]}]
    { \includegraphics[height=1.5in,width=2.2in]{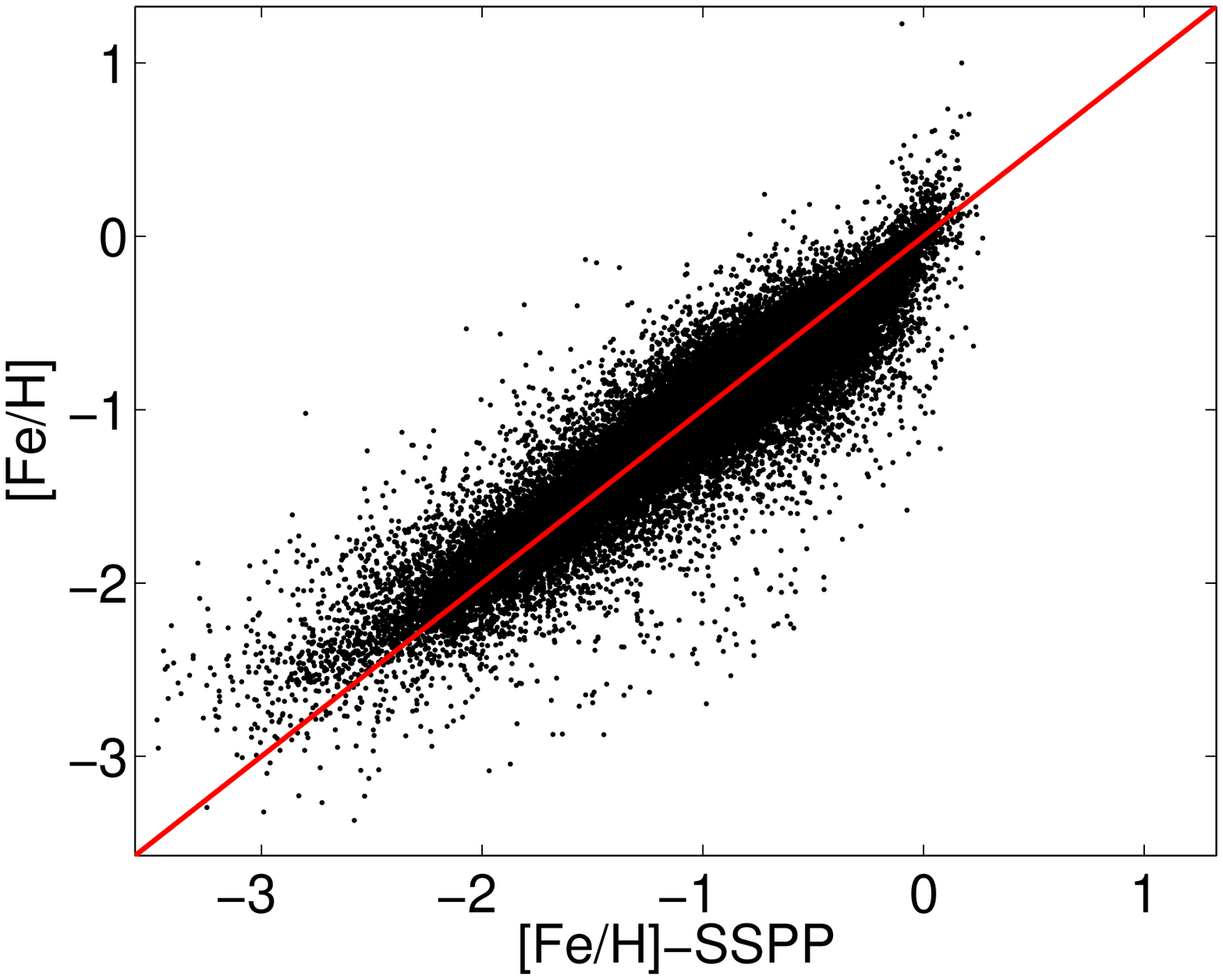} }
  \caption{Performance of the proposed scheme on 40~000 test spectra from SDSS (10~000 SDSS spectra for training, Section \ref{Sec:Data:SDSS}) using SVR$_G$ }
  \label{Fig:comparison:SDSS}
\end{figure*}

\begin{figure*}
  \centering
  \subfigure[log T$_\texttt{eff}$  ]
    { \includegraphics[height=1.5in,width=2.2in]{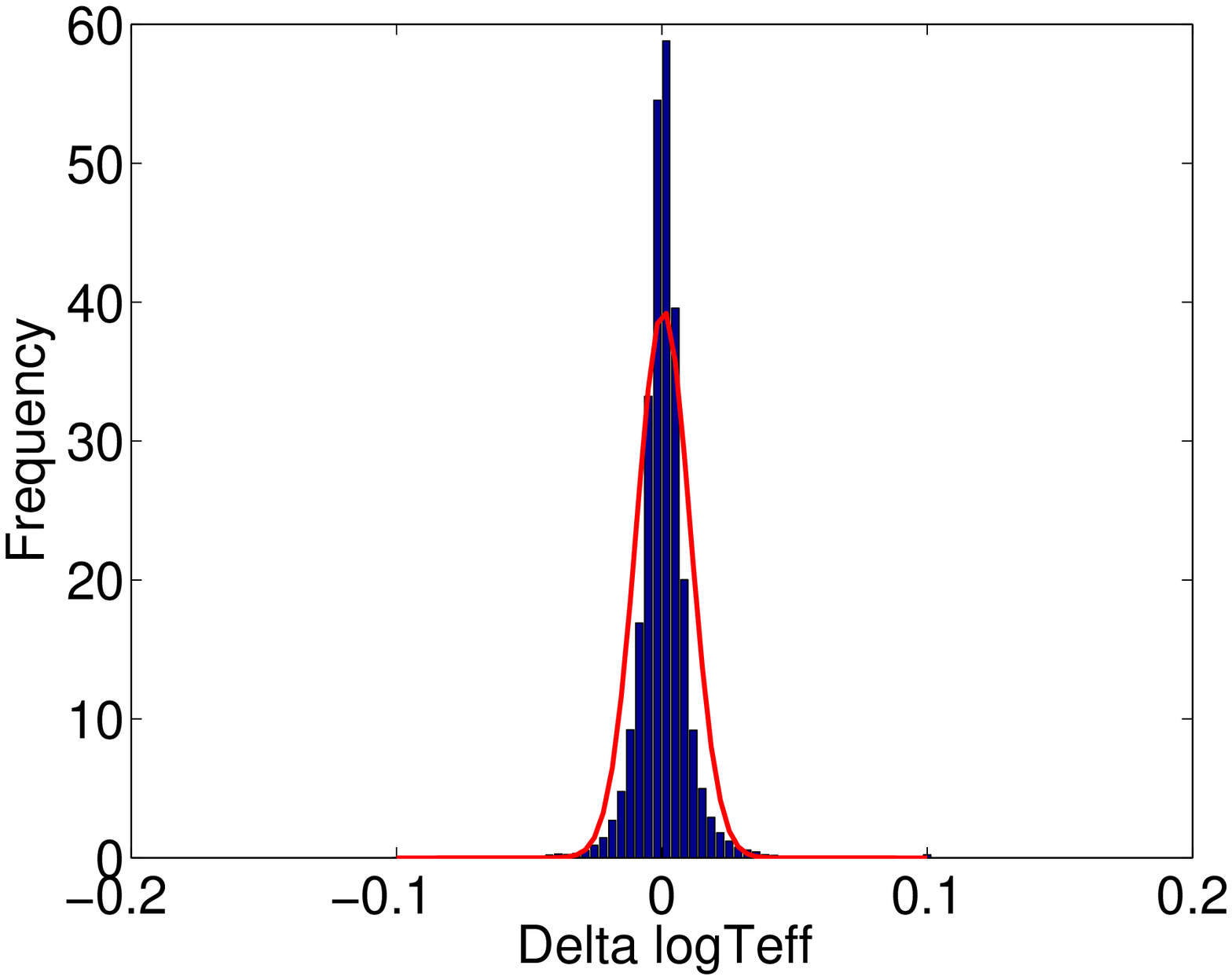} }
  \subfigure[log$~g$]
    { \includegraphics[height=1.5in,width=2.2in]{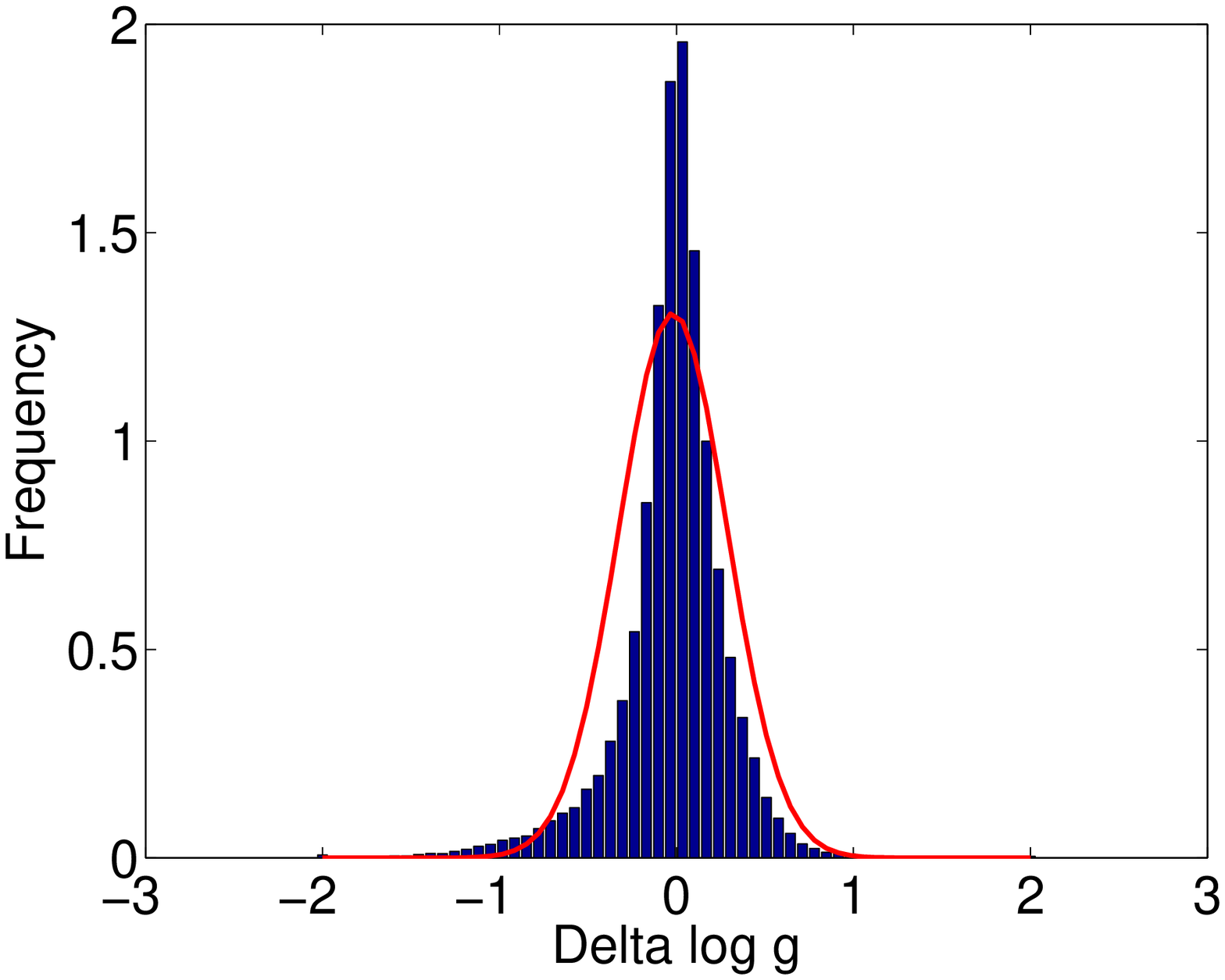} }
  \subfigure[{[Fe/H]}]
    { \includegraphics[height=1.5in,width=2.2in]{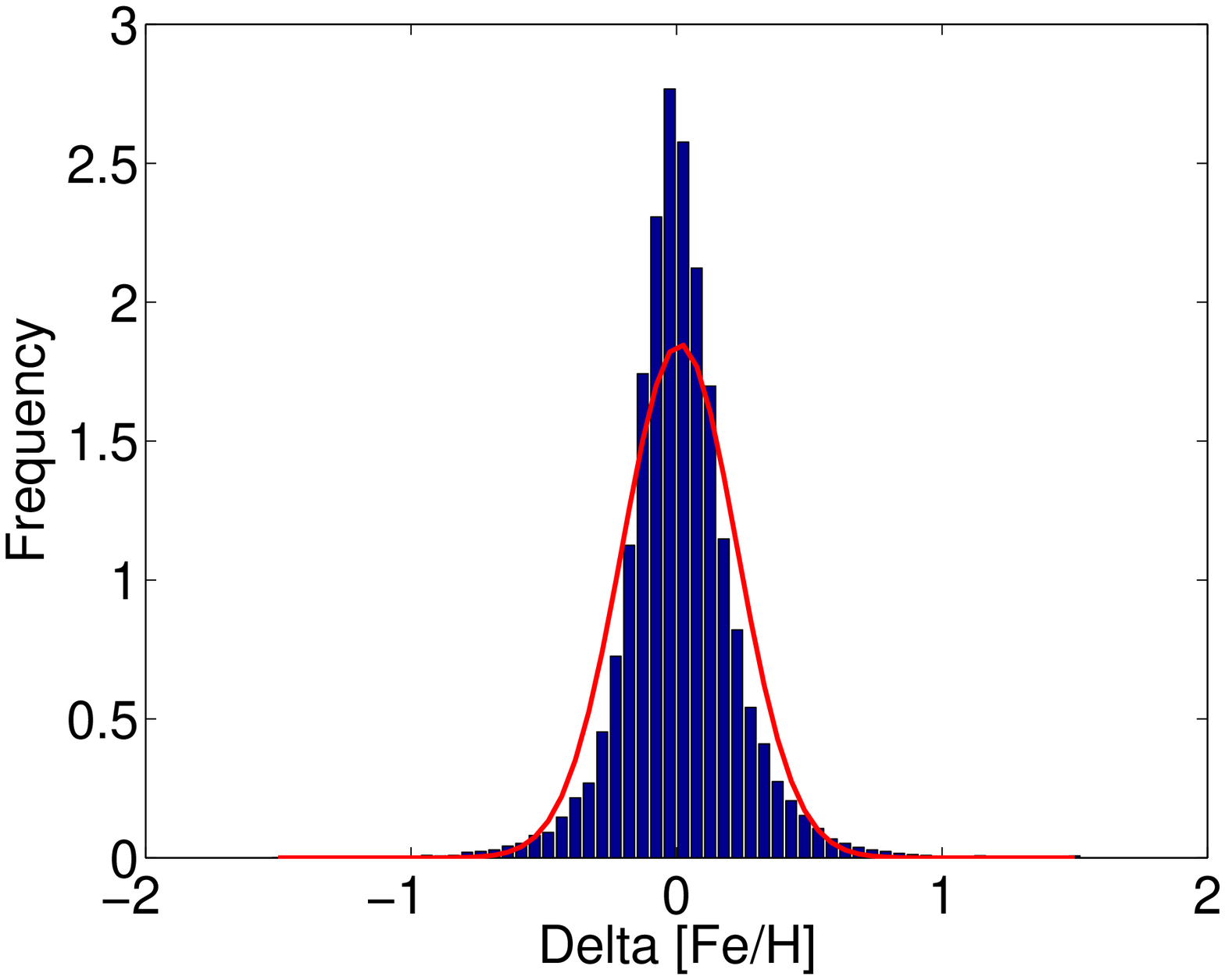} }
  \caption{Residual distributions of the proposed scheme on 40~000 test spectra from SDSS (10~000 SDSS spectra for training, Section \ref{Sec:Data:SDSS})}
  \label{Fig:histogram:SDSS}
\end{figure*}

For the SDSS test set, MAE errors are 0.0062, 0.2035 and 0.1512 dex for log~$T_\texttt{eff}$, log$~g$ and [Fe/H], respectively. To compare the proposed scheme with those in previous related reports, the performance of the proposed scheme was also evaluated using measures ME and SD. More experimental results are presented in Table \ref{table:errors:SDSS} as well as Figs. \ref{Fig:comparison:SDSS} and \ref{Fig:histogram:SDSS}. Direct comparisons with reports are shown in Section \ref{Sec:Conclusion} and more discussion on the dispersion in Fig. \ref{Fig:comparison:SDSS} is presented in Section \ref{Sec:Exper_Disc:Knowlege_Dispersion_Performance}.

\begin{table*}
 \centering
  \caption{Performance of the proposed scheme on 40~000 test spectra from SDSS (10~000 SDSS spectra for training, Section \ref{Sec:Data:SDSS})}
  \begin{tabular}{c|ccc|ccc|ccc}
\hline
   Method &\multicolumn{3}{c}{log T$\texttt{eff}$(T$\texttt{eff}$)} &\multicolumn{3}{c}{log$~g$}  &\multicolumn{3}{c}{[Fe/H]}\\
   Parameter  &MAE &ME &SD   &MAE &ME &SD   &MAE &ME &SD\\
\hline
RBFNN &0.0065(88.48) &4.42$\times 10^{-4}$(6.28)&0.0107(148.04) &0.2159 &0.0205 &0.3228 &0.1547 &6.04$\times 10^{-4}$&0.2197 \\
SVR$_G$ &0.0062(85.83) &6.05$\times 10^{-4}$(9.40)&0.0101(146.66) &0.2035&-0.0193&0.3053 &0.1512 &1.19$\times 10^{-2}$&0.2158 \\
KNNR &0.0069(94.77) &-8.39$\times 10^{-4}$(-10.13)&0.0109(154.62)  &0.2178 &-0.0370&0.3069 &0.2198&-3.56$\times 10^{-2}$ &0.2999 \\
LSR &0.0072(99.22) &3.45$\times 10^{-4}$(5.46)&0.0111(160.73) &0.2594 &0.0270&0.3574  &0.1786 &3.61$\times 10^{-3}$&0.2472\\
SVR$_l$  &0.0070(96.77) &3.41$\times 10^{-4}$(7.11)&0.0111(162.79)  &0.2417 &0.0475&0.3648  &0.1758 &-8.62$\times 10^{-3}$&0.2466\\
\hline
\end{tabular}
Notes. The unit for $T_\texttt{eff}$ is K; The unit for log$T_\texttt{eff}$ is log(K).
\label{table:errors:SDSS}
\end{table*}

\subsection{Performance for LAMOST spectra and synthetic spectra }\label{Sec:Exper_Disc:LAMOST_Synthetic}
The proposed scheme is also tested on actual spectra from LAMOST (section \ref{Sec:Data:LAMOST}) and synthetic spectra (Section \ref{Sec:Data:KURUCZ}).

\begin{figure*}
  \centering
  \subfigure[log T$_\texttt{eff}$  ]
    { \includegraphics[height=1.5in,width=2.2in]{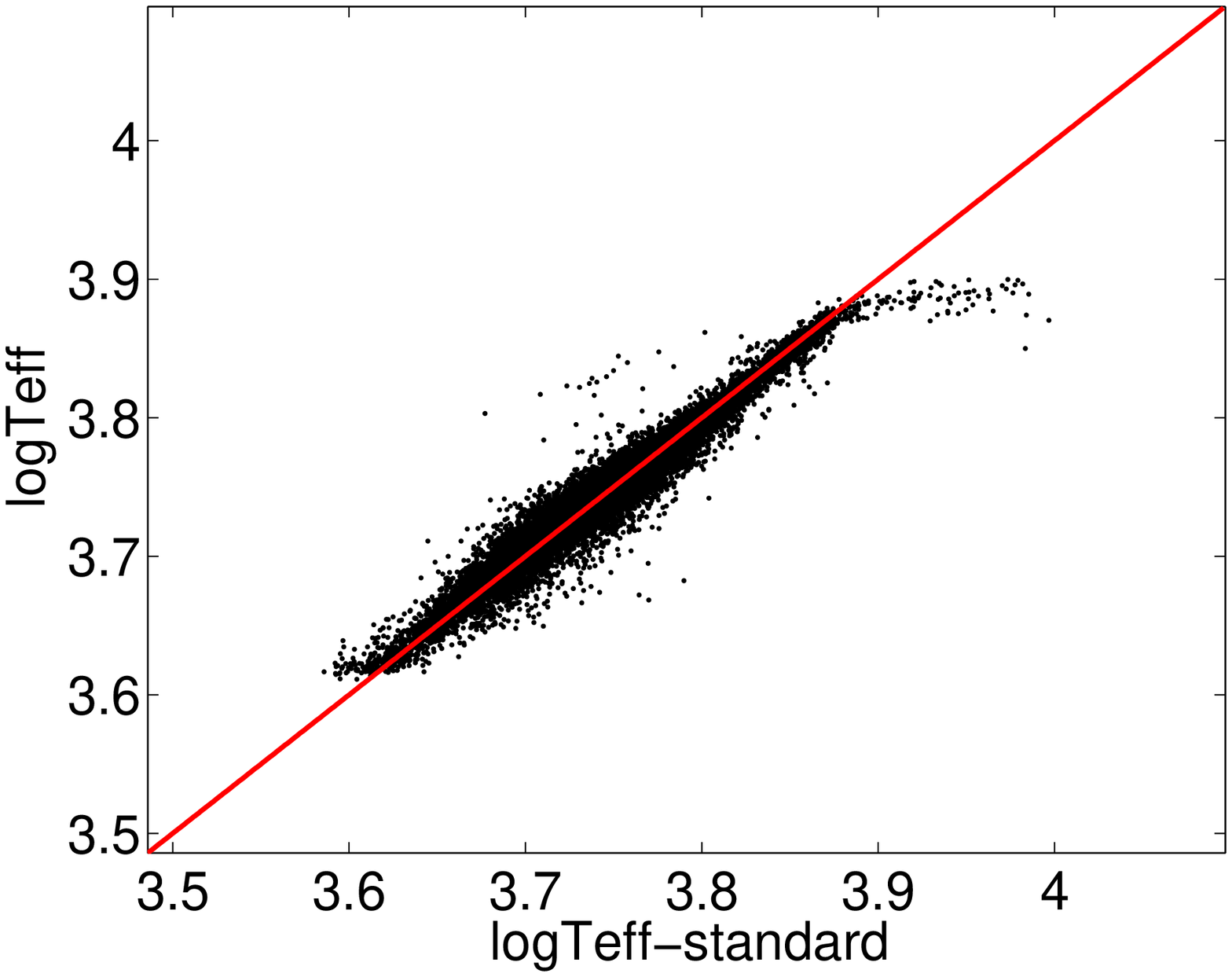} }
  \subfigure[log$~g$]
    { \includegraphics[height=1.5in,width=2.2in]{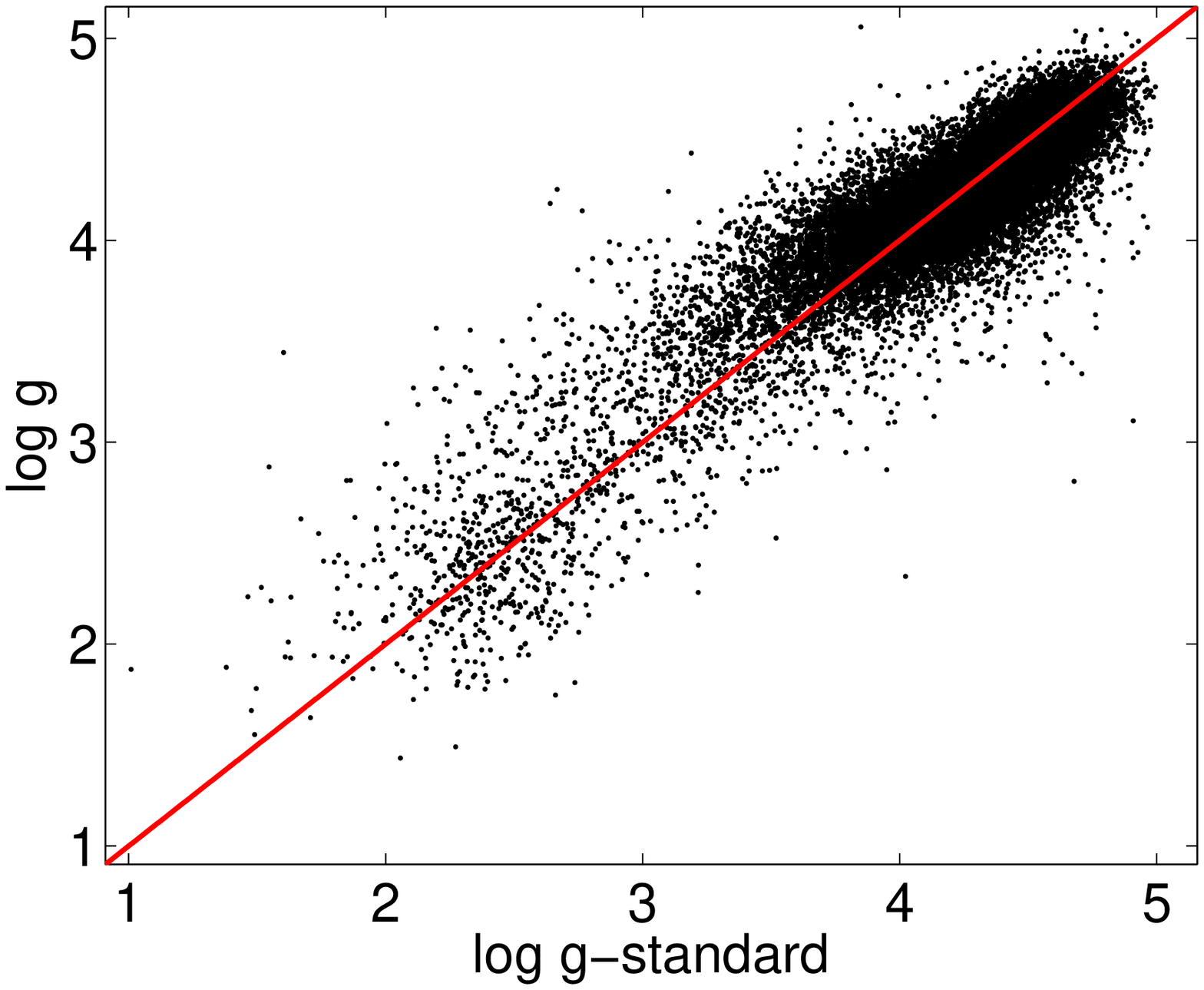} }
  \subfigure[{[Fe/H]}]
    { \includegraphics[height=1.5in,width=2.2in]{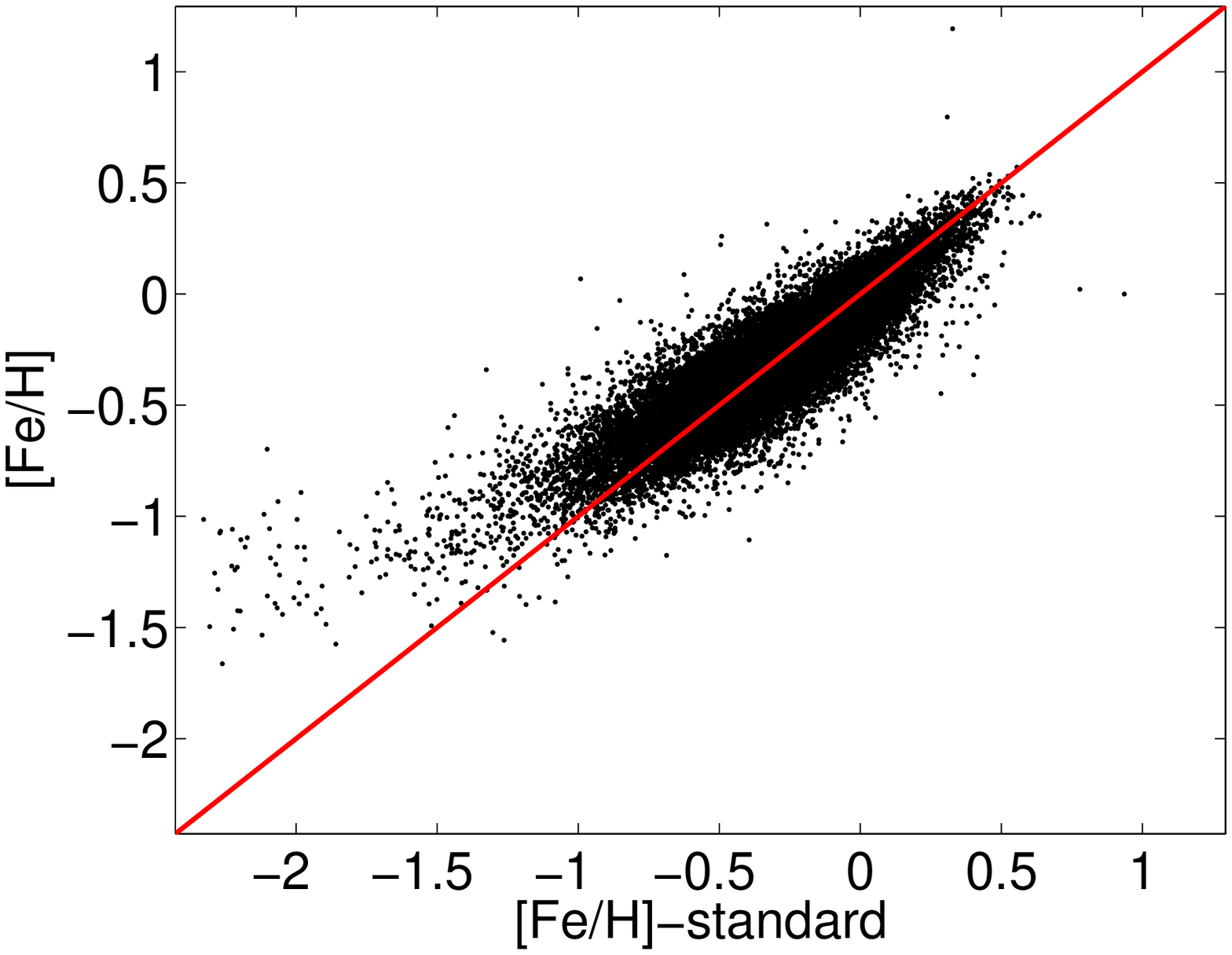} }
  \caption{Performance of the proposed scheme on 23,963 test spectra from LAMOST (10,000 LAMOST spectra for training, Section \ref{Sec:Data:LAMOST}) using SVR$_G$}
  \label{Fig:comparison:LAMOST}
\end{figure*}

\begin{figure*}
  \centering
  \subfigure[log T$_\texttt{eff}$  ]
    { \includegraphics[height=1.5in,width=2.2in]{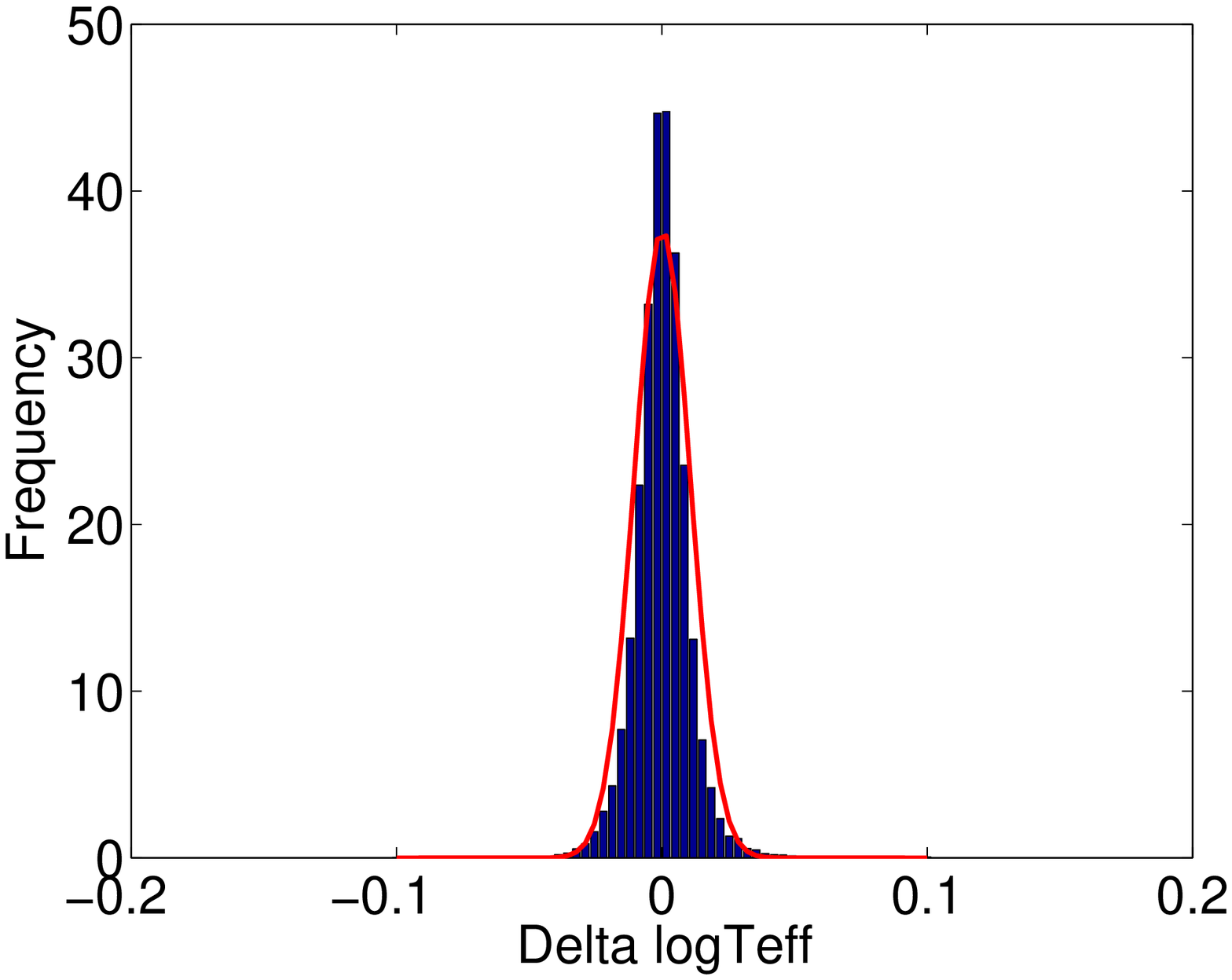} }
  \subfigure[log$~g$]
    { \includegraphics[height=1.5in,width=2.2in]{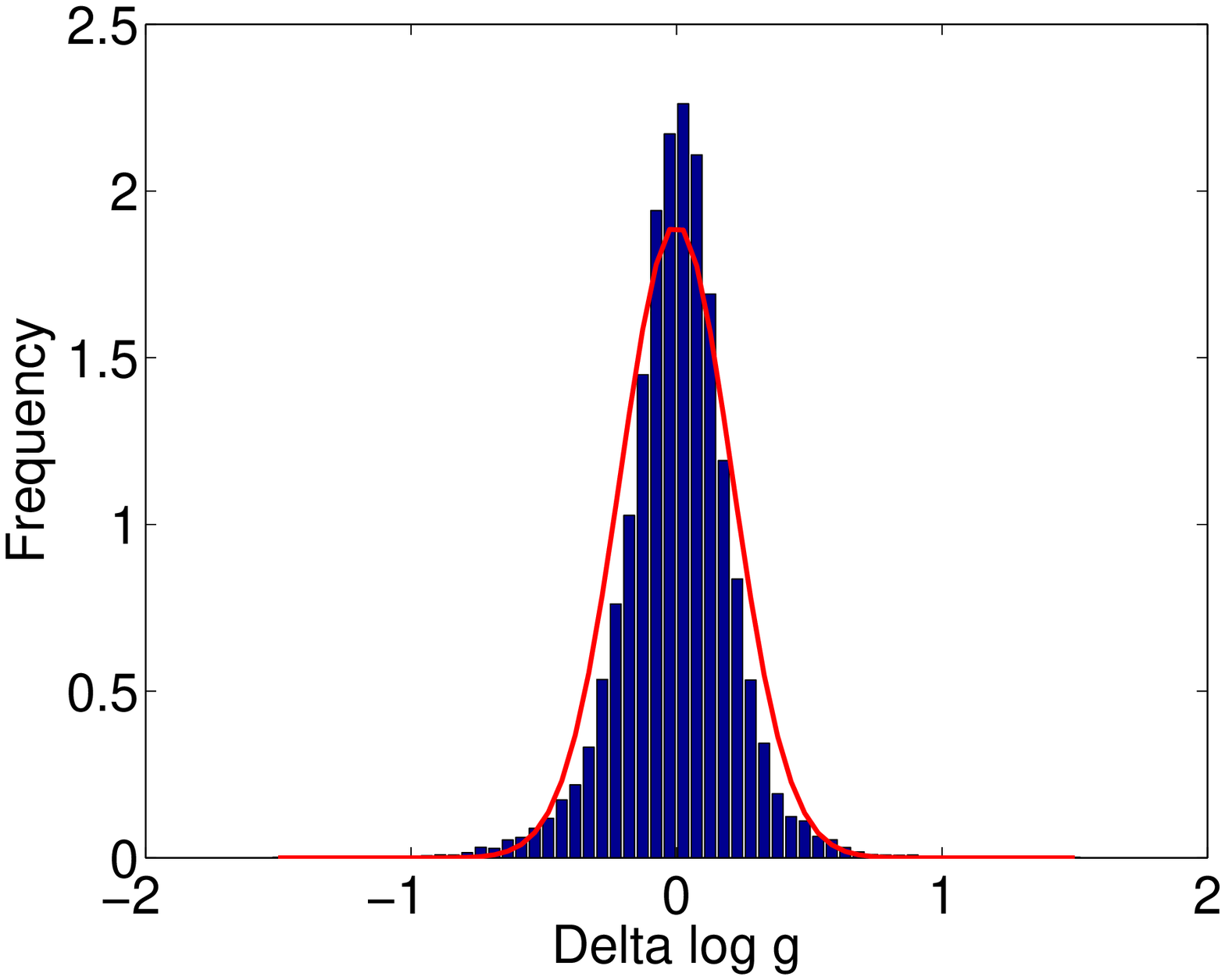} }
  \subfigure[{[Fe/H]}]
    { \includegraphics[height=1.5in,width=2.2in]{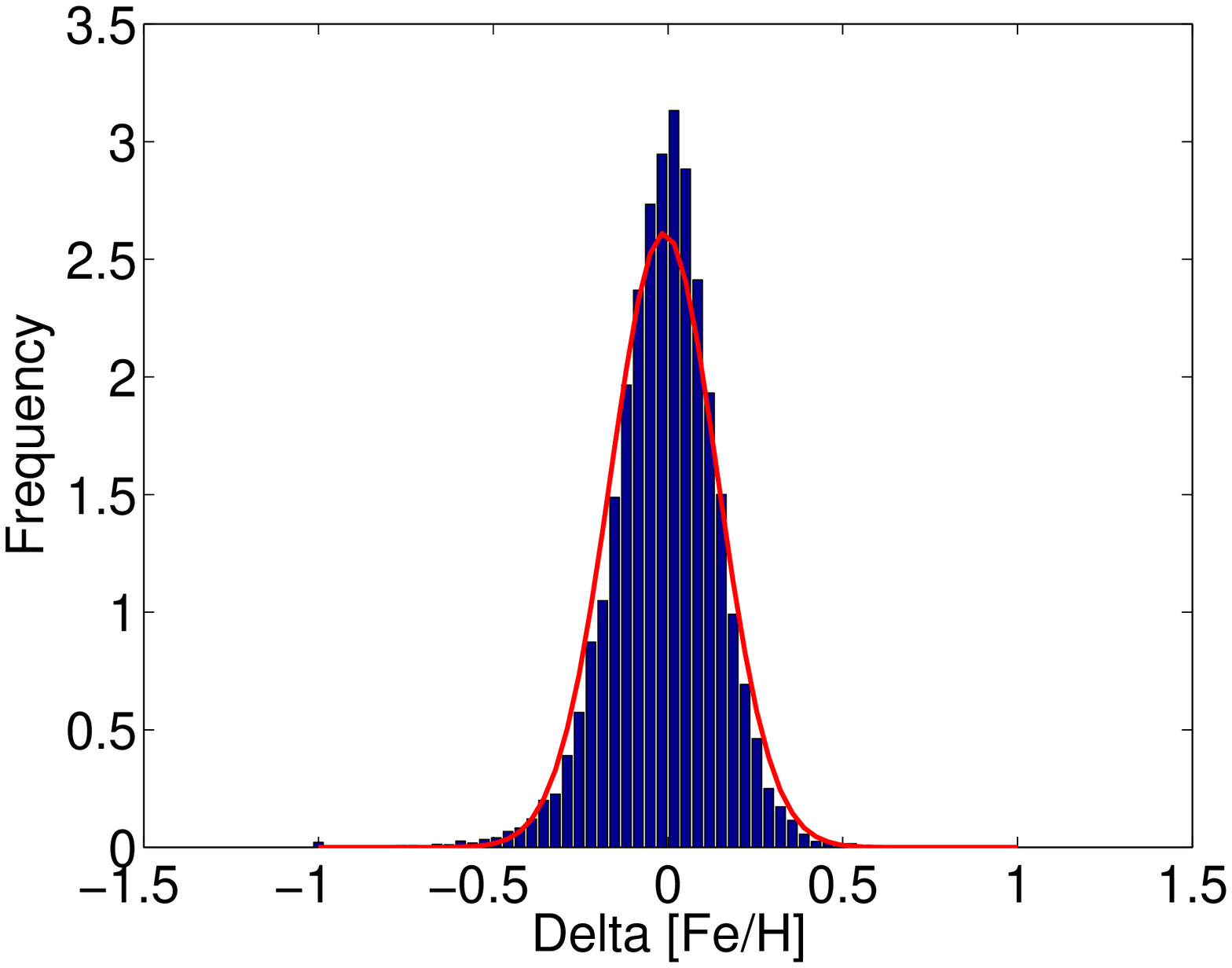} }
  \caption{Residual distributions of the proposed scheme on 23,963 test spectra from LAMOST (10000 LAMOST spectra for training, Section \ref{Sec:Data:LAMOST}) using SVR$_G$}
  \label{Fig:histogram:LAMOST}
\end{figure*}

\begin{figure*}
  \centering
  \subfigure[log T$_\texttt{eff}$  ]
    { \includegraphics[height=1.5in,width=2.2in]{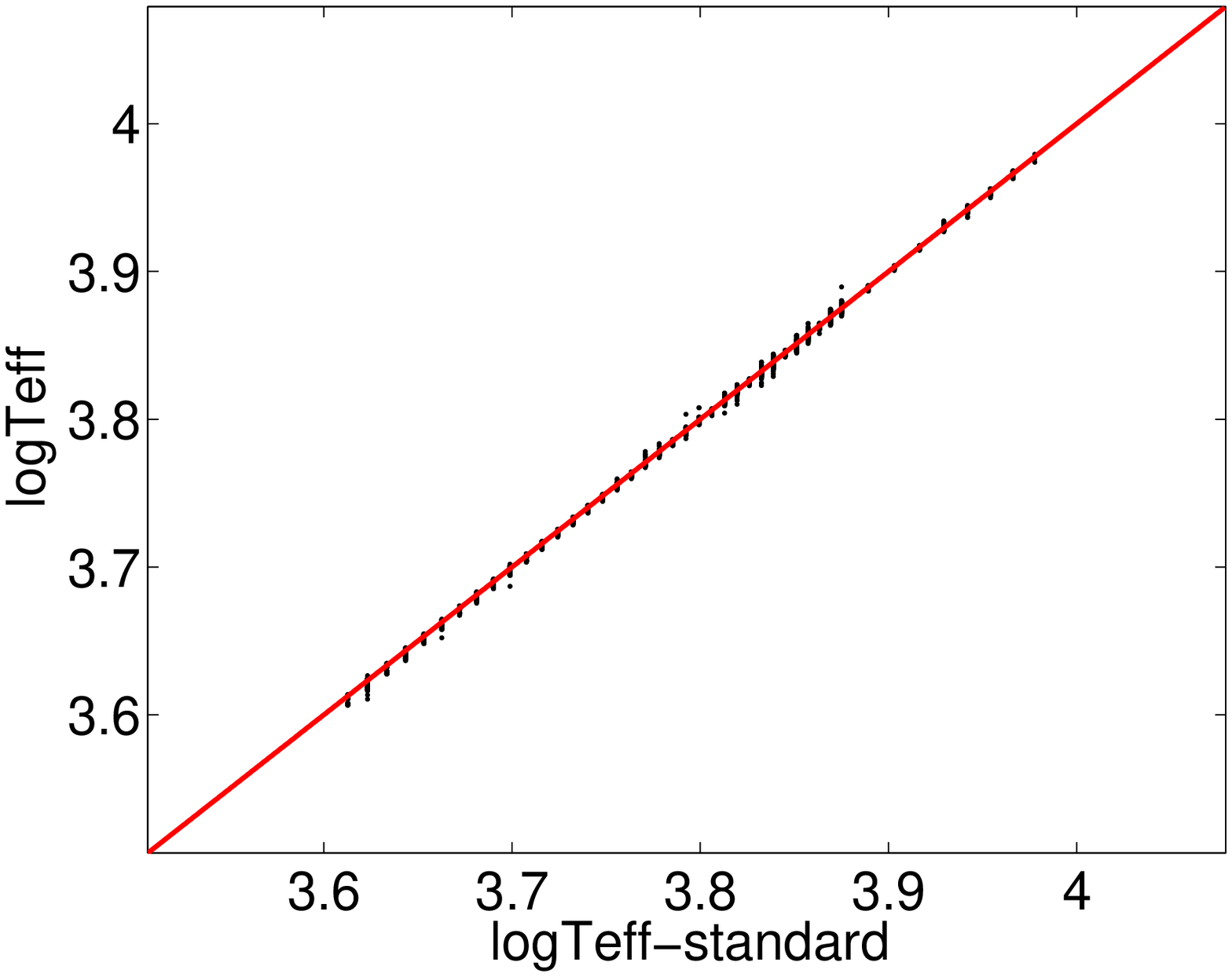} }
  \subfigure[log$~g$]
    { \includegraphics[height=1.5in,width=2.2in]{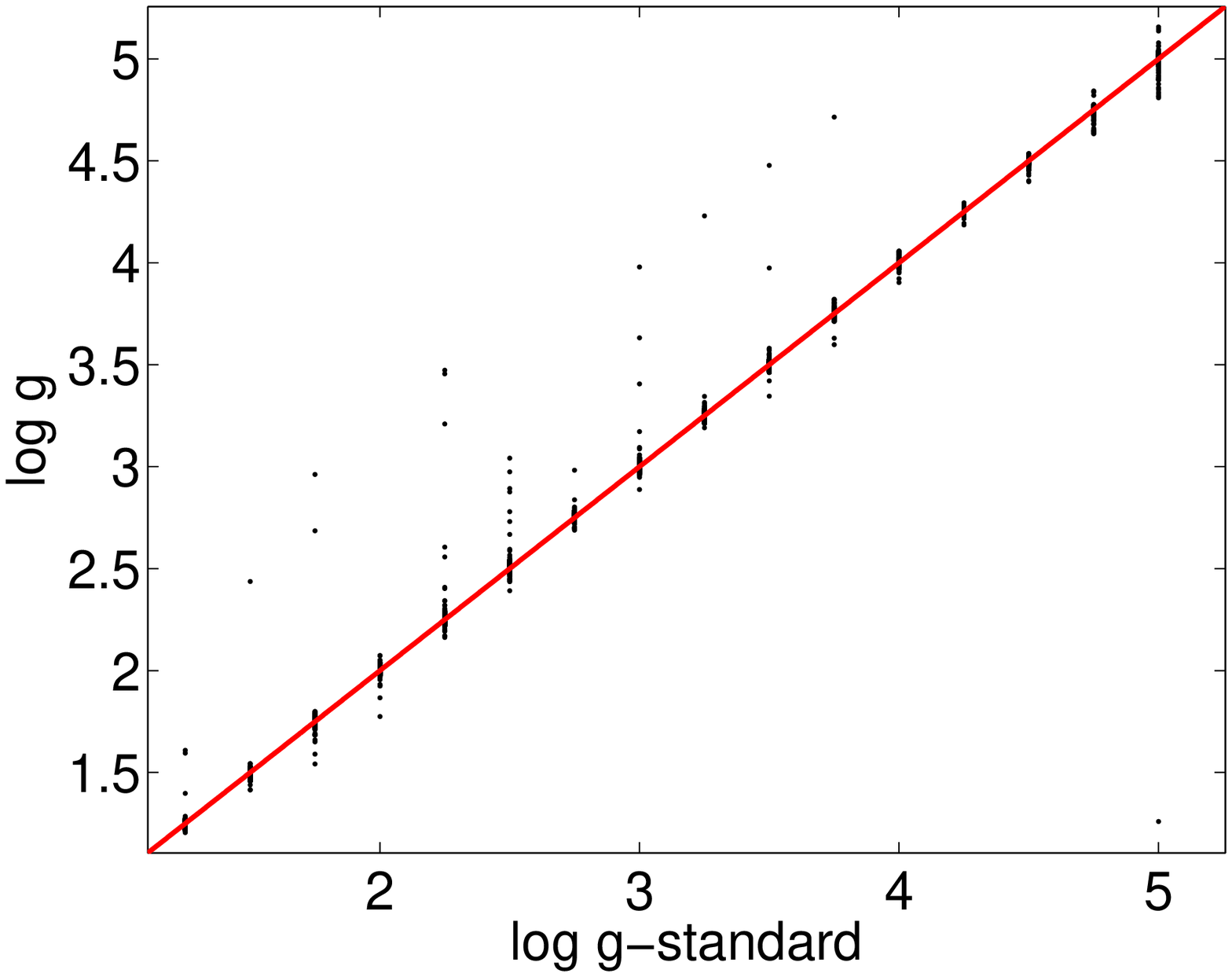} }
  \subfigure[{[Fe/H]}]
    { \includegraphics[height=1.5in,width=2.2in]{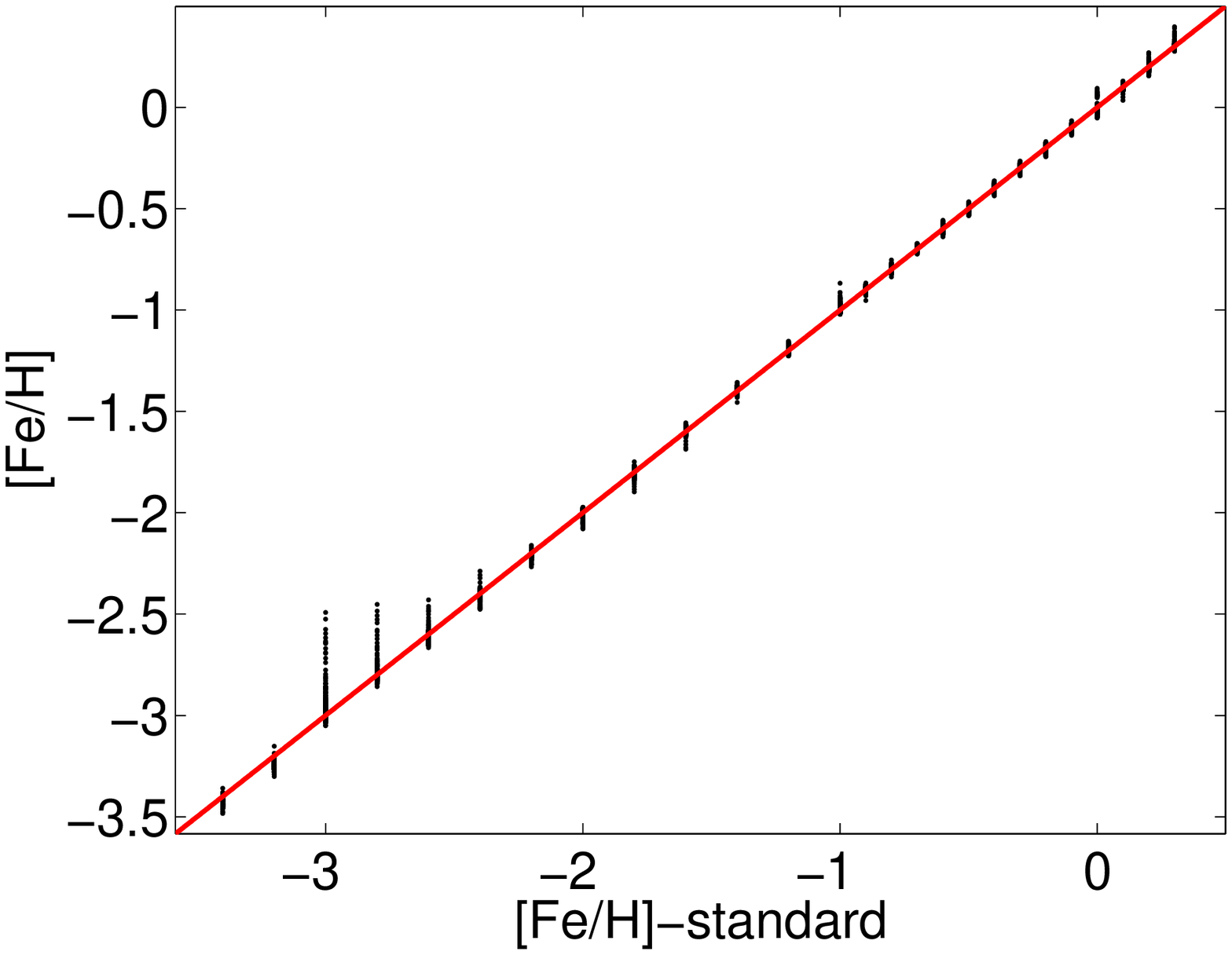} }
  \caption{Performance of the proposed scheme on 10,469 test spectra computed from KURUCZ's model (8,500 synthetic spectra for training, Section \ref{Sec:Data:KURUCZ}) using SVR$_G$}
  \label{Fig:comparison:KURUCZ:8500}
\end{figure*}

\begin{figure*}
  \centering
  \subfigure[log T$_\texttt{eff}$  ]
    { \includegraphics[height=1.5in,width=2.2in]{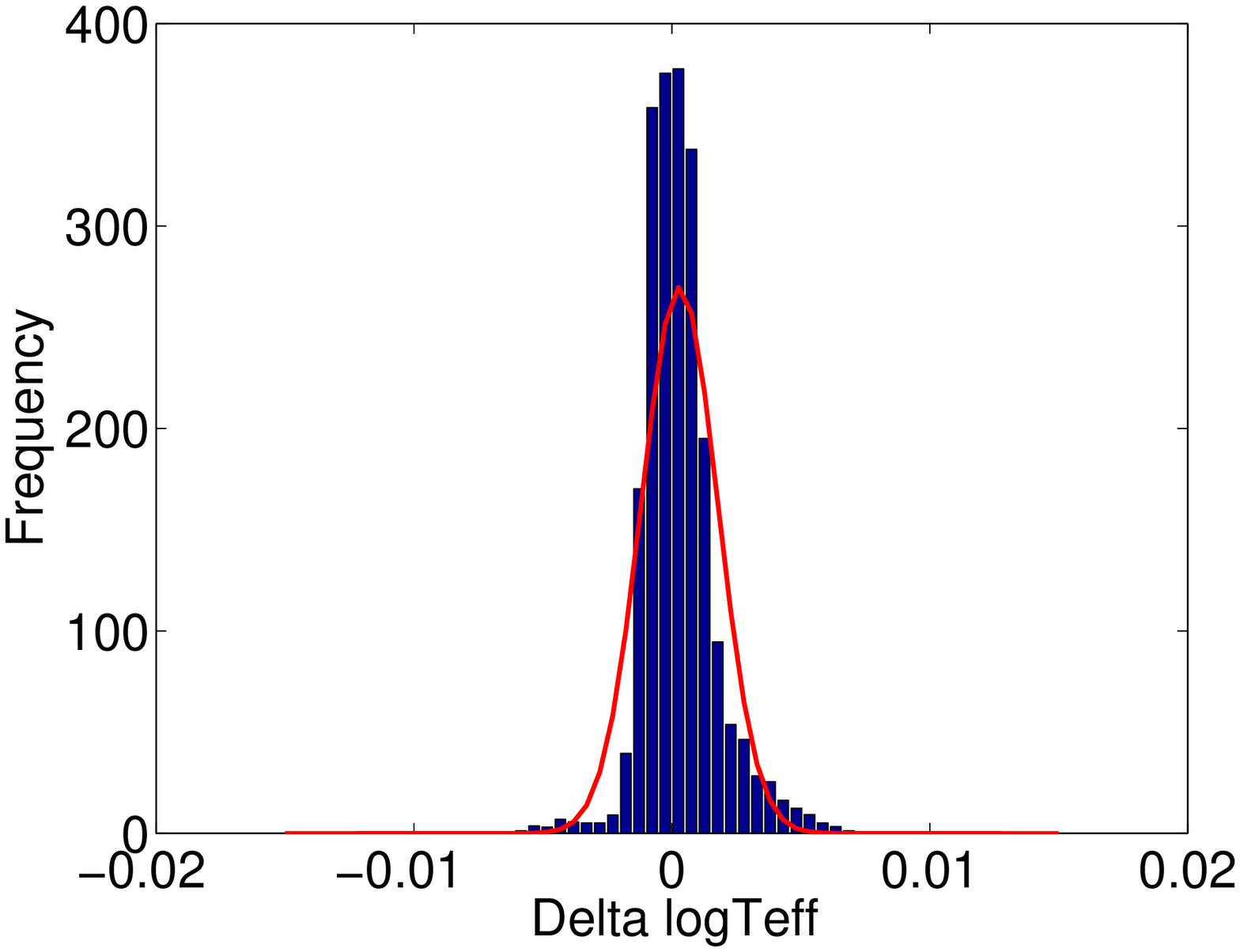} }
  \subfigure[log$~g$]
    { \includegraphics[height=1.5in,width=2.2in]{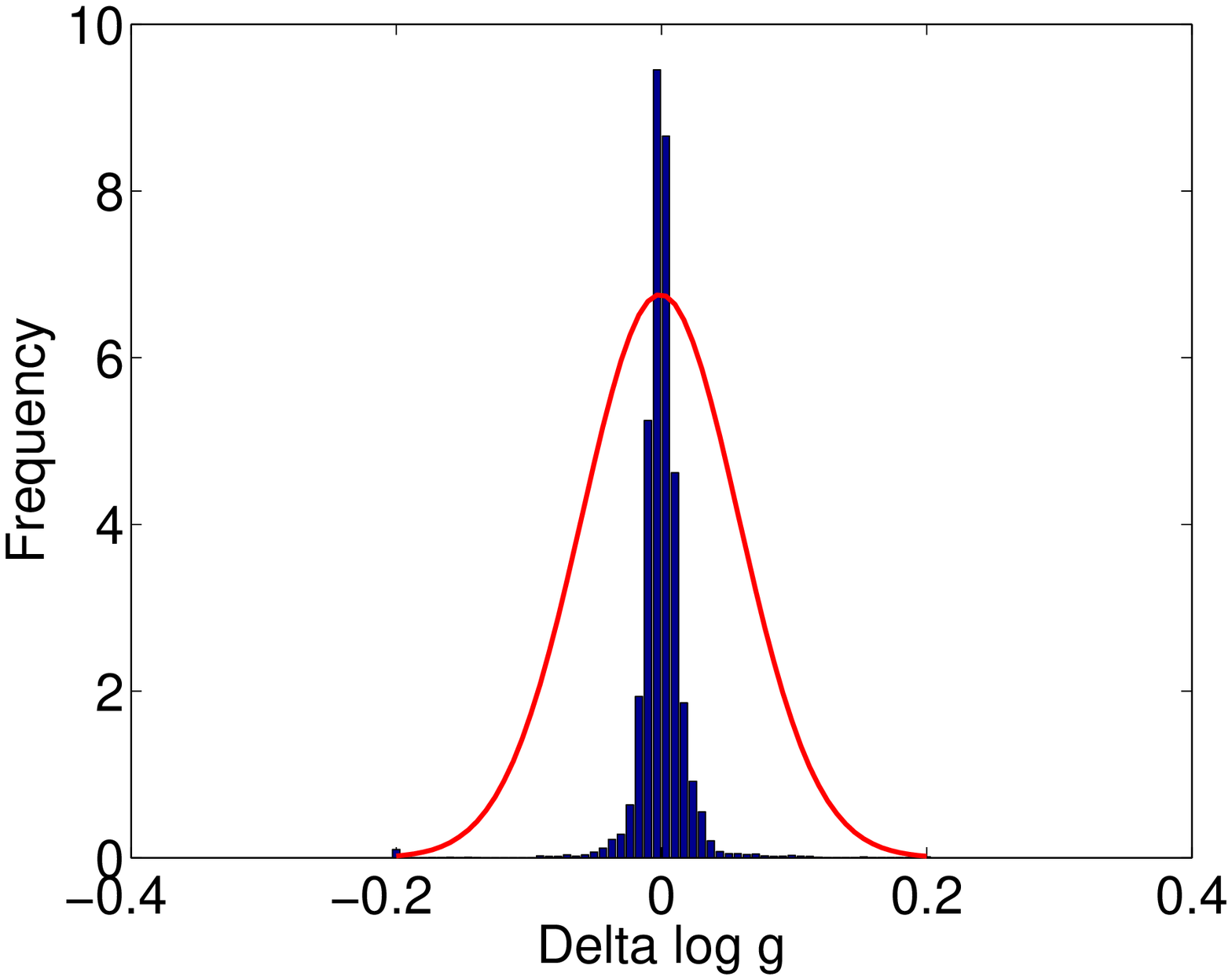} }
  \subfigure[{[Fe/H]}]
    { \includegraphics[height=1.5in,width=2.2in]{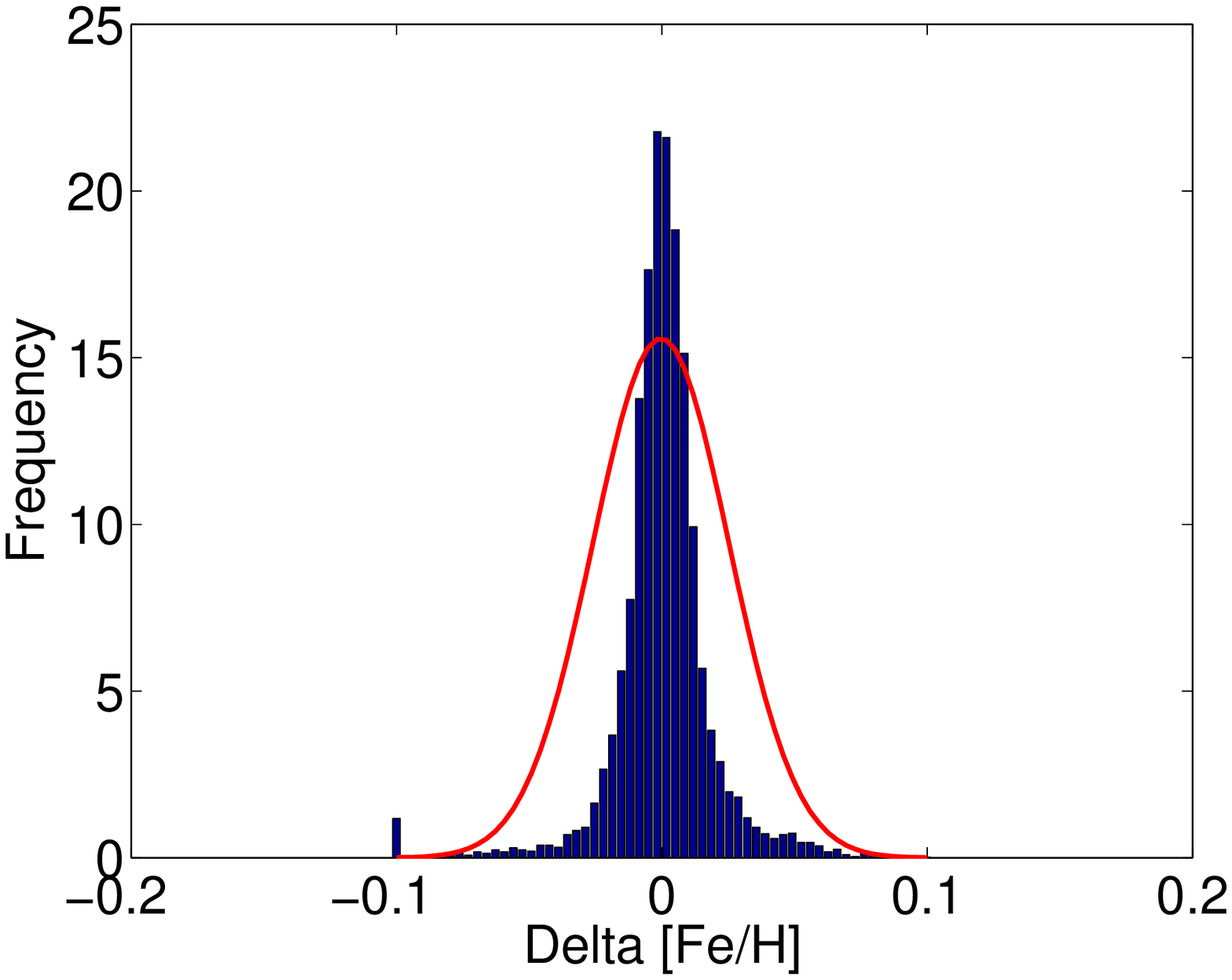} }
  \caption{Residual distributions of the proposed scheme on 10,469 test spectra computed from KURUCZ's model (8,500 synthetic spectra for training, Section \ref{Sec:Data:KURUCZ}) using SVR$_G$}
  \label{Fig:histogram:KURUCZ}
\end{figure*}

The performances of the scheme on LAMOST spectra is presented in Table \ref{table:errors:Lamost}, as well as Figs. \ref{Fig:comparison:LAMOST} and \ref{Fig:histogram:LAMOST}. The test results using synthetic spectra are presented in Table \ref{table:errors:KURUCZ}, as well as Figs. \ref{Fig:comparison:KURUCZ:8500} and \ref{Fig:histogram:KURUCZ}. The results in Tables \ref{table:errors:SDSS}, \ref{table:errors:Lamost} and \ref{table:errors:KURUCZ} show that the SVR$_G$ and RBFNN are more suitable than KNNR, LSR and SVR$_l$ for estimating atmospheric parameters.

\begin{table*}
 \centering
  \caption{Performance of the proposed scheme on 23~963 test spectra from LAMOST (10~000 LAMOST spectra for training, Section \ref{Sec:Data:LAMOST})}
  \begin{tabular}{c|ccc|ccc|ccc}
  \hline
   Method &\multicolumn{3}{c}{log T$\texttt{eff}$(T$\texttt{eff}$)} &\multicolumn{3}{c}{log$~g$}  &\multicolumn{3}{c}{[Fe/H]}\\
   Parameter  &MAE &ME &SD   &MAE &ME &SD   &MAE &ME &SD\\
\hline
  RBFNN  &0.0070(91.14) &6.75$\times 10^{-5}$(2.37)&0.0099(131.36)   &0.1664 &0.0109 &0.2753     &0.1197 &-0.0038 &0.1767\\
  SVR$_G$ &0.0074(95.37)&1.27$\times 10^{-4}$(4.30)&0.0106(141.62)  &0.1528 &-0.0008 &0.2102   &0.1146 &-0.0112 &0.1528\\
  KNNR &0.0085(111.80)&4.47$\times 10^{-4}$(10.08)&0.0126(173.66)  &0.1934 &0.0167 &0.2730   &0.1625 &-0.0154 &0.2151\\
  LSR &0.0082(106.51)&6.67$\times 10^{-4}$(11.46)&0.0117(161.83)  &0.2218 &0.0173 &0.3404  &0.1312 &-0.0143 &0.1807\\
  SVR$_l$ &0.0081(105.73)&1.89$\times 10^{-4}$(5.44)&0.0116(159.83)  &0.2070&-0.0124&0.3145  &0.1311&-0.0222 &0.1806\\
\hline
\end{tabular}
Notes. The unit for $T_\texttt{eff}$ is K; The unit for log$T_\texttt{eff}$ is log(K).
\label{table:errors:Lamost}
\end{table*}

\begin{table*}
 \centering
  \caption{Performance of the proposed scheme on 10~469 test spectra computed from KURUCZ's model (8~500 synthetic spectra for training, Section \ref{Sec:Data:KURUCZ})}
  \begin{tabular}{c|ccc|ccc|ccc}
  \hline
   Method &\multicolumn{3}{c}{log T$\texttt{eff}$(T$\texttt{eff}$)} &\multicolumn{3}{c}{log$~g$}  &\multicolumn{3}{c}{[Fe/H]}\\
   Parameter  &MAE &ME &SD   &MAE &ME &SD   &MAE &ME &SD\\
\hline
RBFNN &0.0010 (14.15)&1.34$\times 10^{-4}$(1.42)&0.0014(20.26) &0.0217 &1.89$\times 10^{-3}$&0.0582 &0.0203 &1.95$\times 10^{-3}$&0.0282\\
SVR$_G$&0.0010 (14.42)&2.63$\times 10^{-4}$(3.34)&0.0015(20.81) &0.0123 &-9.42$\times 10^{-4}$&0.0590 &0.0125 &-3.58$\times 10^{-4}$&0.0256 \\
KNNR &0.0027(39.39) &3.92$\times 10^{-5}$(0.19)&0.0041 (61.08)&0.2167 &3.55$\times 10^{-2}$&0.3166 &0.1007 &2.90$\times 10^{-2}$&0.1611 \\
LSR &0.0026 (36.18)&-2.54$\times 10^{-4}$(-3.68)&0.0033(46.05) &0.1416 &2.36$\times 10^{-2}$&0.1902 &0.0903 &9.77$\times 10^{-3}$&0.1175  \\
SVR$_l$ &0.0025 (34.91)&1.14$\times 10^{-4}$(1.79)&0.0032 (45.80)&0.1343 &2.21$\times 10^{-2}$&0.1920 &0.0783 &4.12$\times 10^{-3}$&0.1122\\
\hline
\end{tabular}
Notes. The unit for $T_\texttt{eff}$ is K; The unit for log$T_\texttt{eff}$ is log(K).
\label{table:errors:KURUCZ}
\end{table*}

\subsection{Filtering and Selection - Positive or Negative?}\label{Sec:Exper_Disc:Filtering_Selection}
The proposed scheme extracts spectral features by removing high-frequency components from the Haar wavelet transform and rejecting most low-frequency components by the LASSO algorithm. In this study, we determine whether these processes eliminate important spectral information, such as weak lines.

Four experiments are conducted and the results are listed in Table \ref{table:Compactness}. The results show the possibility of eliminating useful spectral information. However, in application, the observed spectrum is inevitably contaminated with noise. In theory, weak lines should be more sensitive to noise.

Therefore, the loss of the elimination is trivial. The wavelet components with the lowest frequency are traditional choices for spectral features for estimating atmospheric parameters \citep{Lu2013}. In the experiments for T$_\texttt{eff}$, when all low-frequency wavelet components are used, the number of features will increase from 17 to 239 (increase (239-17)/17 = 1305.88 percent), but the MAE can only decrease by 0.0007 dex (11.29 percent: Experiment 1 and 2). When no component is eliminated while estimating T$_\texttt{eff}$, the number of features will increase from 17 to 3823 (an increase of 22388.23 percent), but the MAE error increases by 0.0398 dex (641.93 percent). A small number of detected features indicates an efficient process for estimating atmospheric parameters from spectral features. The above results suggest that the model developed in this work can estimate stellar atmospheric parameters with high accuracy.

\begin{table*}
 \centering
  \caption{In these experiments, the advantages and disadvantages of eliminating high-frequency, as well as many low-frequency, components are evaluated. The parameters are estimated by support vector machine (RBF kernel) and RBF neural network on SDSS samples. Their performances are assessed by MAE. WT(i,0) and WT(i,1) represent the coefficients of a wavelet transform with $i-$level decomposition on the approximation and high-frequency sub-bands. $\{T_{i}\}$, $\{L_{i}\}$, and $\{F_{i}\}$ denote the extracted features for log T$_\texttt{eff}$, log$~g$, and [Fe/H], respectively. The number behind ``:" represents the total number of the utilised features.
   }
  \begin{tabular}{c|ccc|ccc|ccc|}
  \hline
  \multicolumn{3}{c}{log T$_\texttt{eff}$}  &\multicolumn{3}{c}{log$~g$} &\multicolumn{3}{c}{[Fe/H]}  \\
  Features    &SVR$_G$  &RBFNN  &features  &SVR  &RBFNN &features   &SVR$_G$  &RBFNN\\
 \hline
  $\{T_{i}\}$:17 &0.0062 &0.0065 &$\{L_{i}\}$:24 &0.2035 &0.2159 &$\{F_{i}\}$:25 &0.1512 &0.1547\\
 \hline
  WT(4,0):239  &0.0055  &0.0062 &WT(4,0):239  &0.1909 &0.2267 &WT(4,0):239  &0.1311 &0.1486\\
 \hline
  WT(4,1)+$\{T_{i}\}$:256 &0.0165 &0.0083 &WT(4,1)+$\{L_{i}\}$:263 &0.2368 &0.2449 &WT(4,1)+$\{F_{i}\}$:264 &0.1862 &0.1770\\
\hline
  Full:3823  &0.0460 &0.0131 &Full:3823   &0.3726 &0.2366  & Full:3823    &0.4118 &0.1769\\
\hline
\end{tabular}
Notes. The unit for $T_\texttt{eff}$ is K; The unit for log$T_\texttt{eff}$ is log(K).
\label{table:Compactness}
\end{table*}

\subsection{Knowledge base, dispersion and performance}\label{Sec:Exper_Disc:Knowlege_Dispersion_Performance}
The proposed scheme is a statistical learning method. Its primary principle is to discover automatically the mapping from a stellar spectrum to its atmospheric parameters from a training set. The training set is the carrier of knowledge and affects the accuracy of the scheme.

\begin{figure*}
  \centering
  \subfigure[learn from 15000 SDSS spectra for log T$_\texttt{eff}$  ]
    { \includegraphics[height=1.5in,width=2.2in]{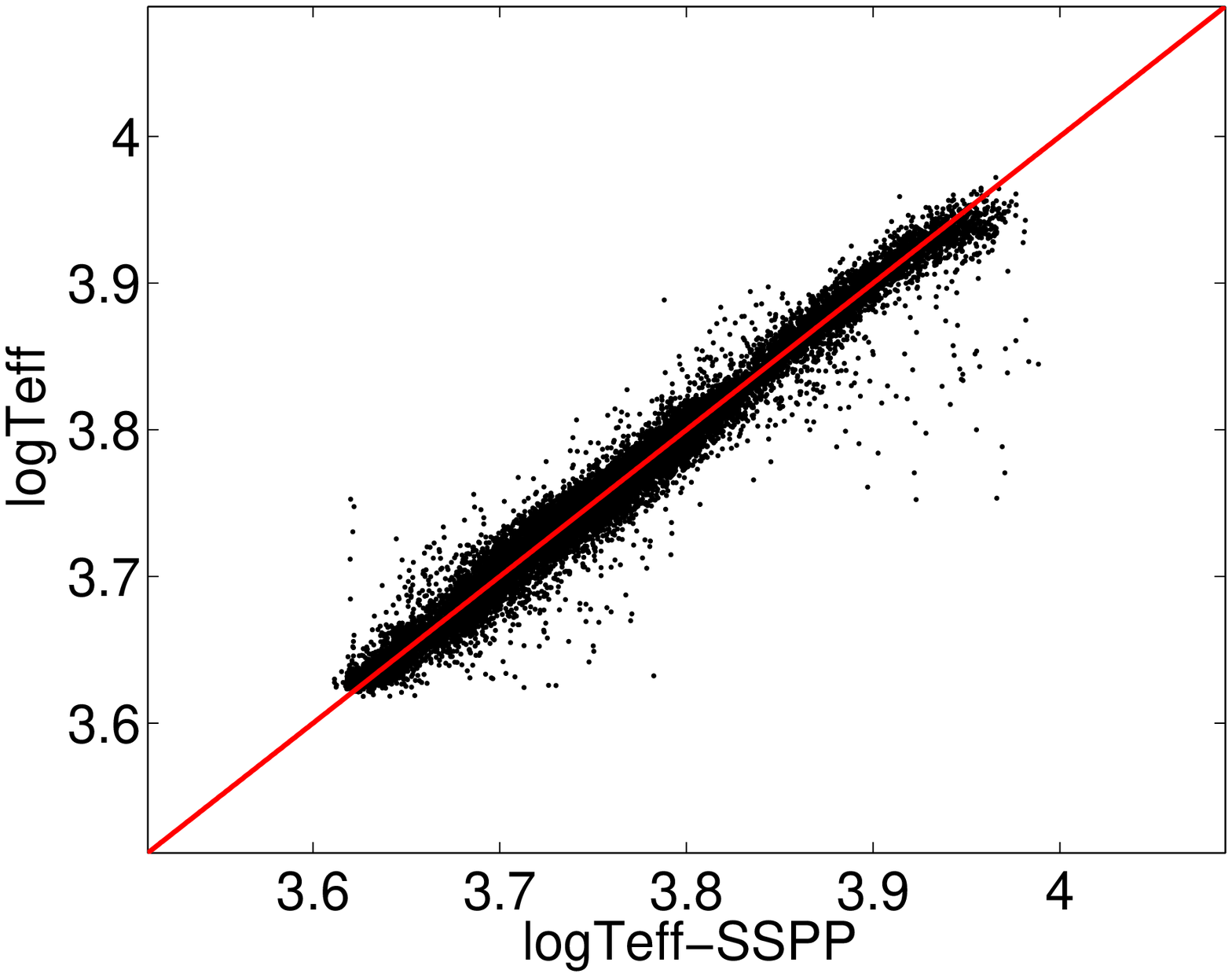} }
  \subfigure[learn from 15000 SDSS spectra for log$~g$]
    { \includegraphics[height=1.5in,width=2.2in]{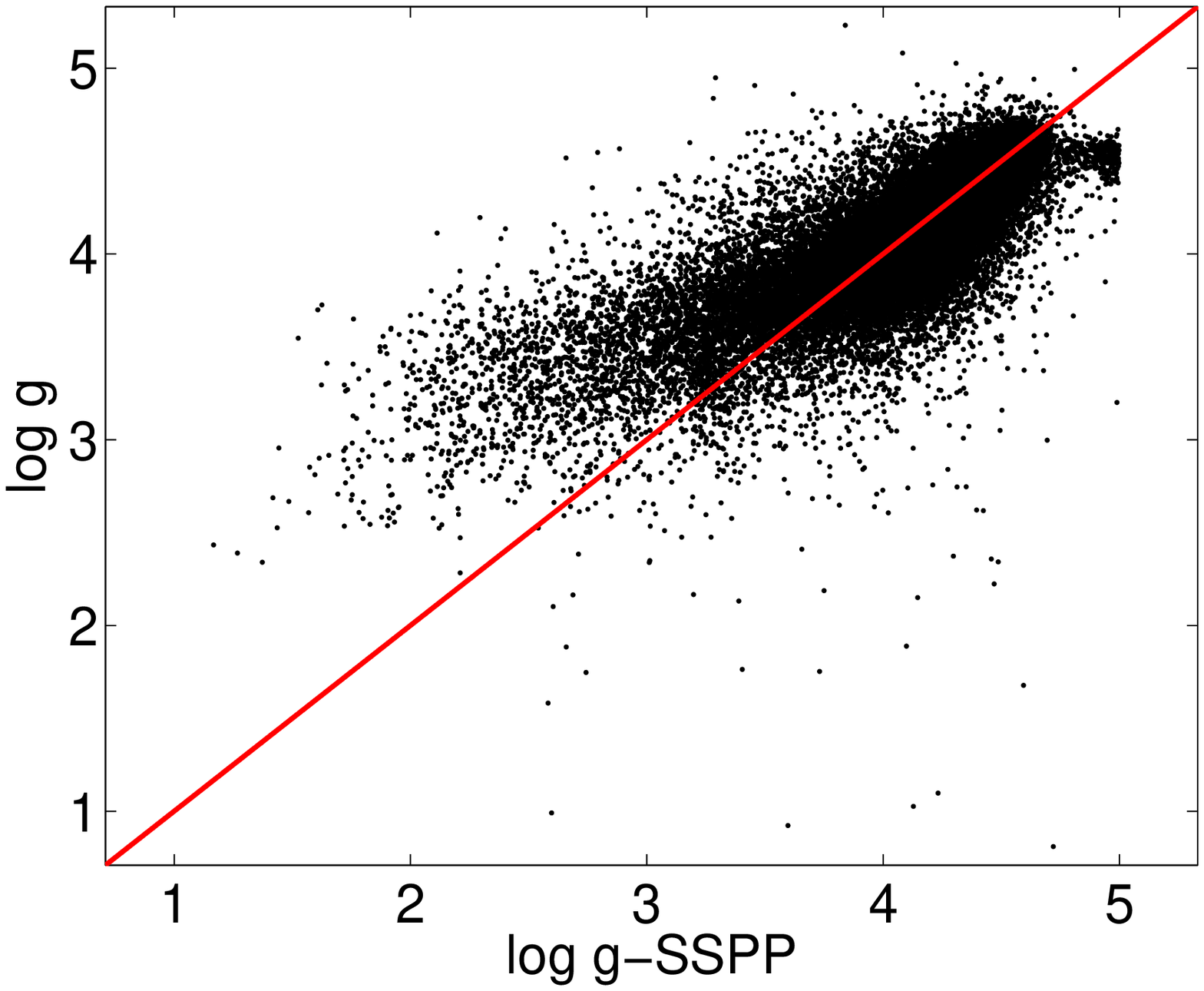} }
  \subfigure[learn from 15000 SDSS spectra for {[Fe/H]}]
    { \includegraphics[height=1.5in,width=2.2in]{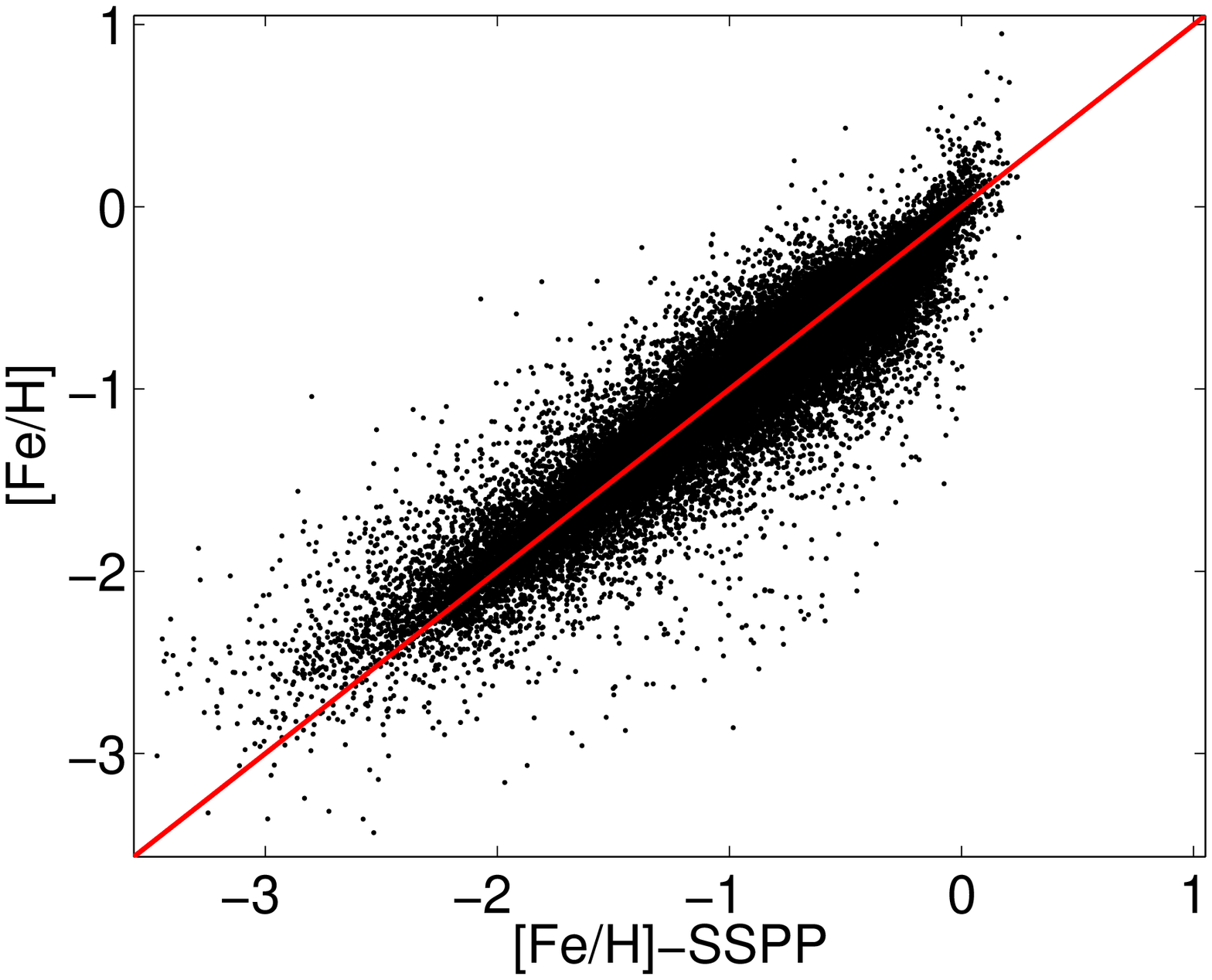} }
  \subfigure[learn from 20000 SDSS spectra for log T$_\texttt{eff}$  ]
    { \includegraphics[height=1.5in,width=2.2in]{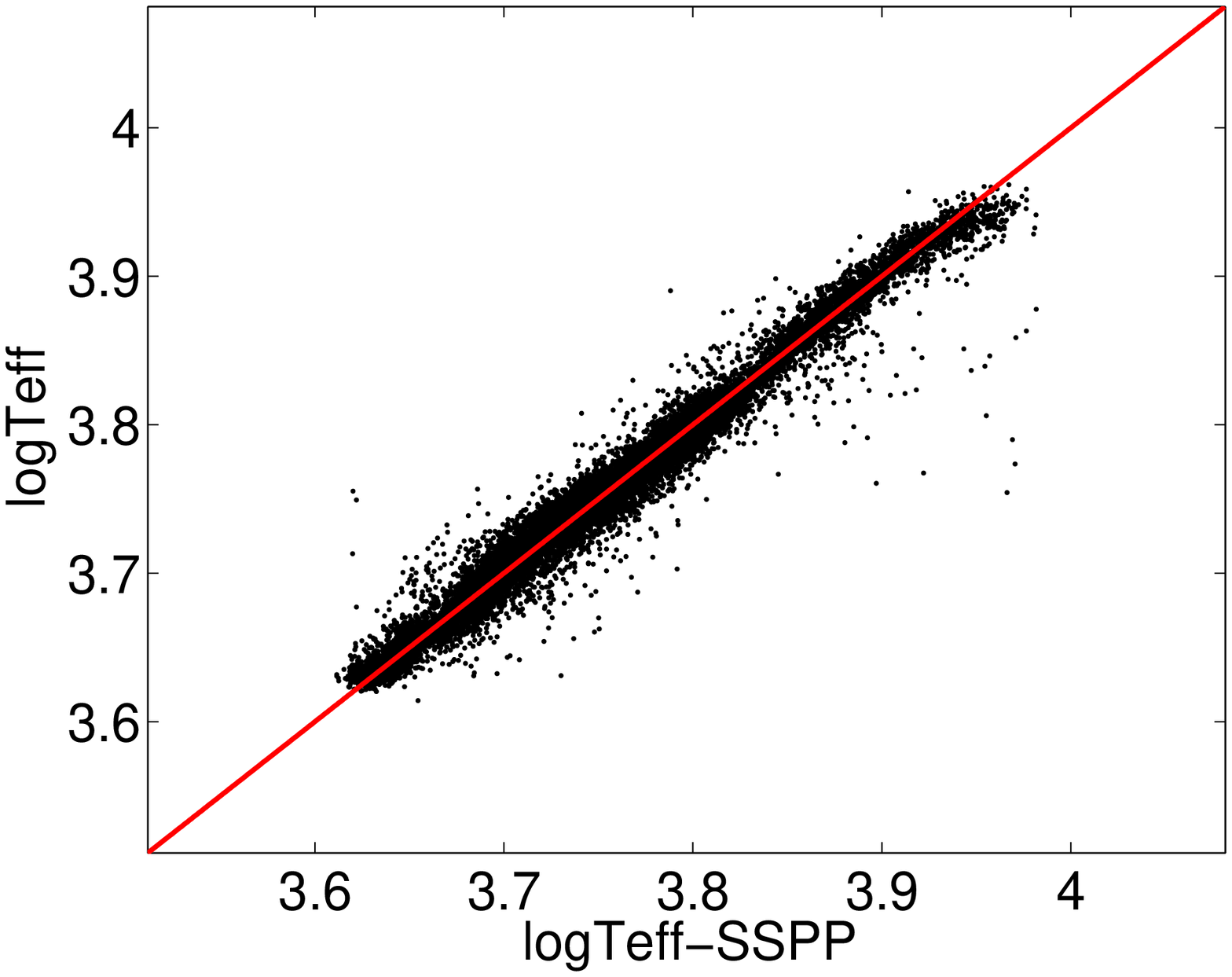} }
  \subfigure[learn from 20000 SDSS spectra for log$~g$]
    { \includegraphics[height=1.5in,width=2.2in]{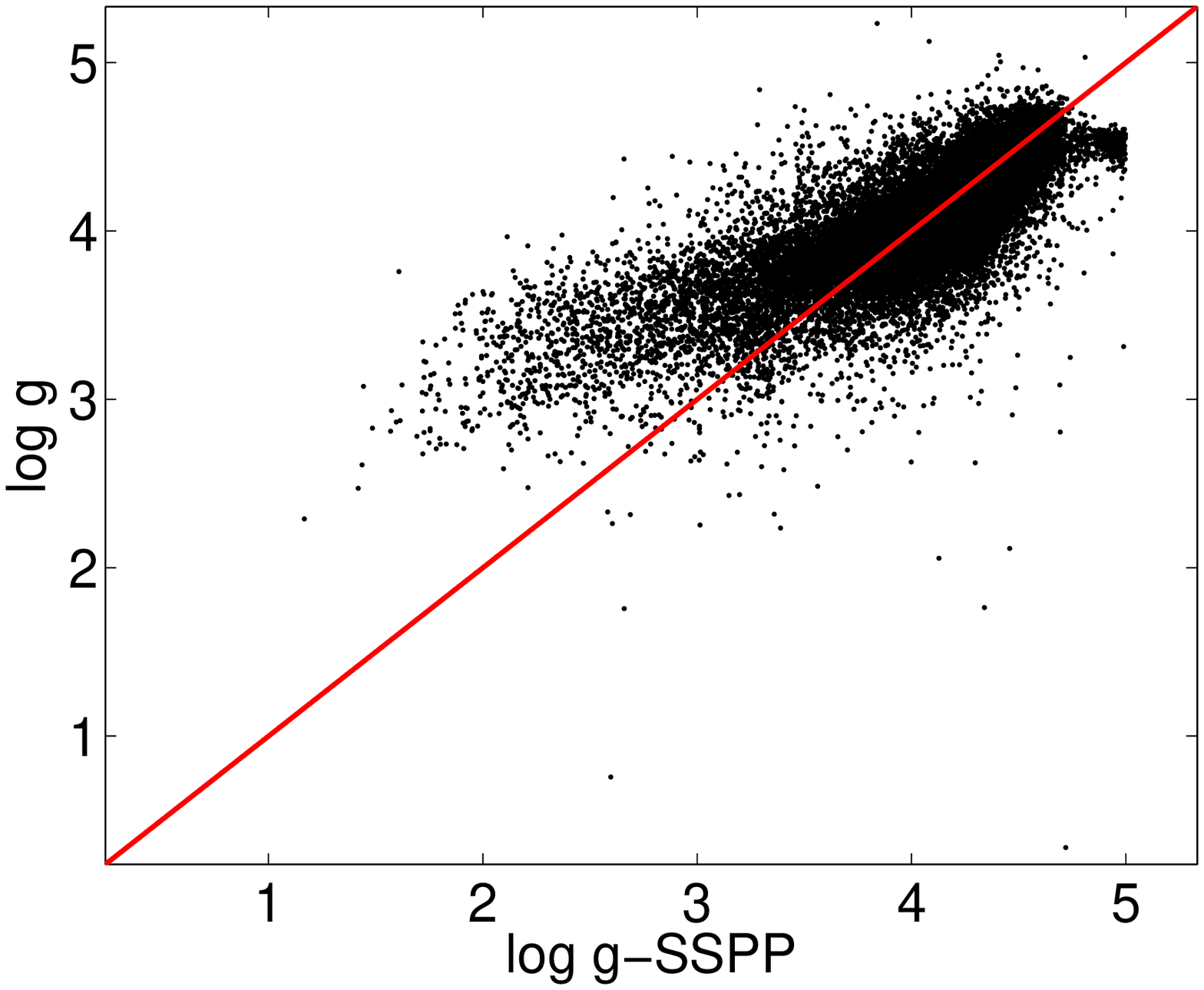} }
  \subfigure[learn from 20000 SDSS spectra for {[Fe/H]}]
    { \includegraphics[height=1.5in,width=2.2in]{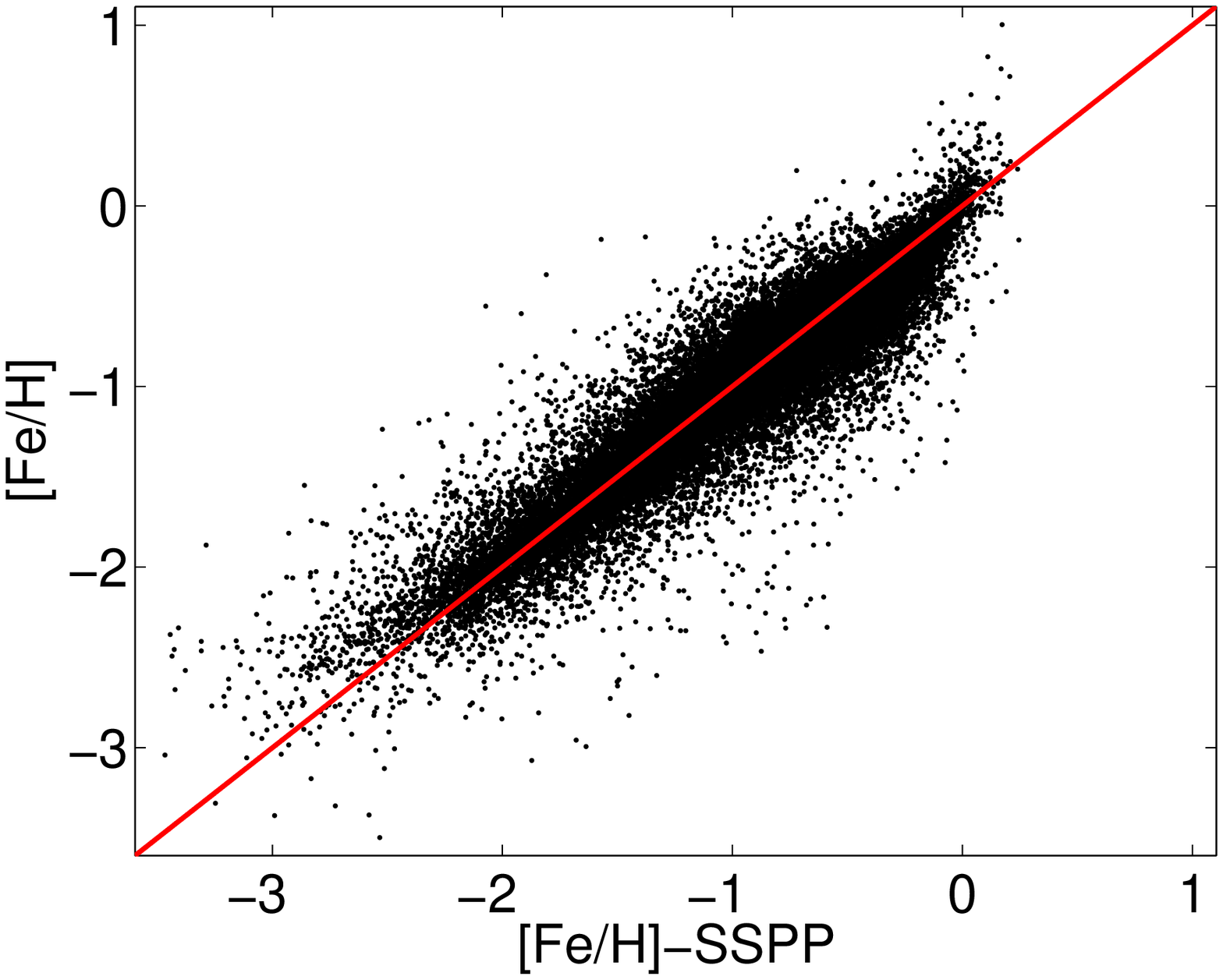} }
  \subfigure[learn from 25000 SDSS spectra for log T$_\texttt{eff}$  ]
    { \includegraphics[height=1.5in,width=2.2in]{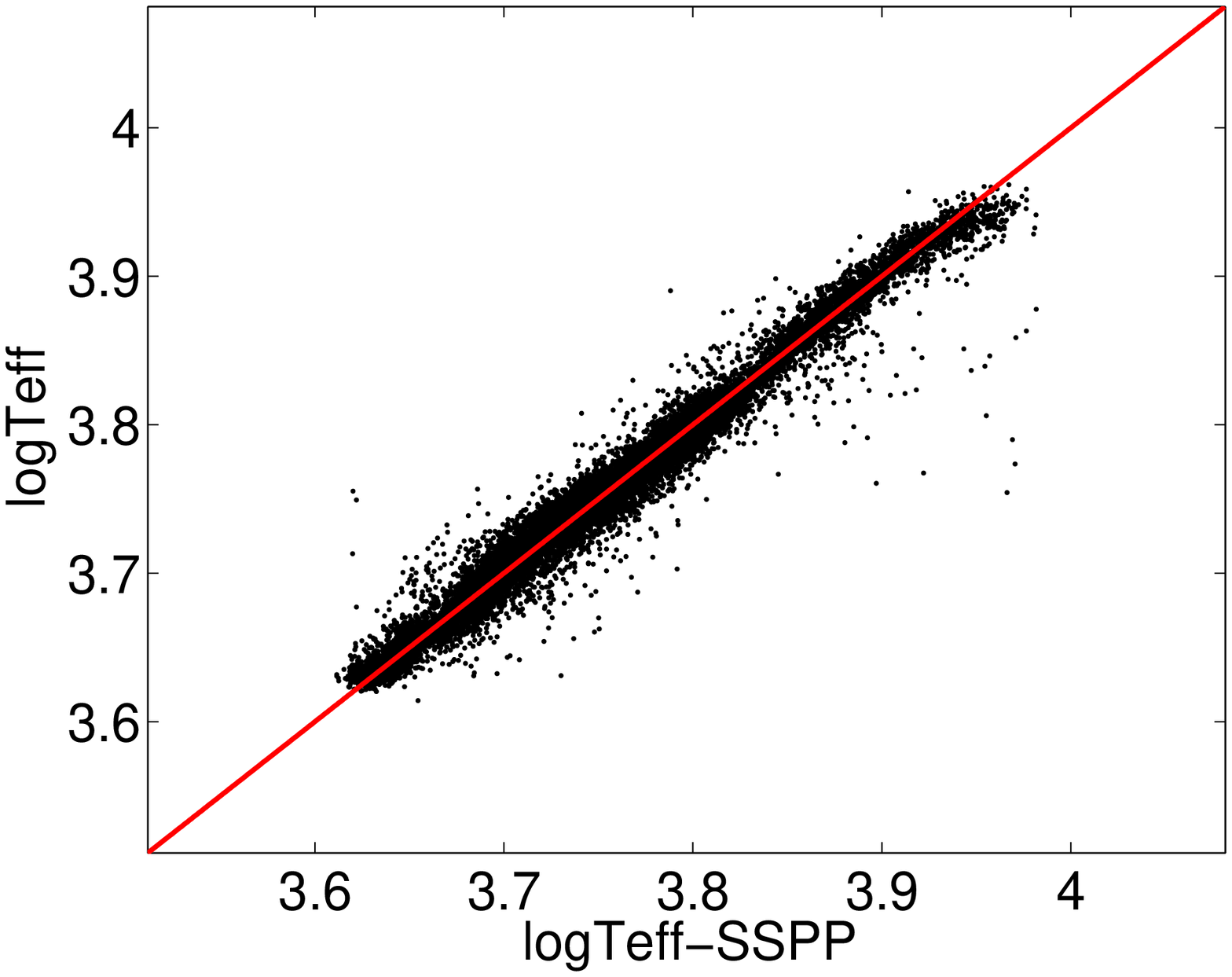} }
  \subfigure[learn from 25000 SDSS spectra for log$~g$]
    { \includegraphics[height=1.5in,width=2.2in]{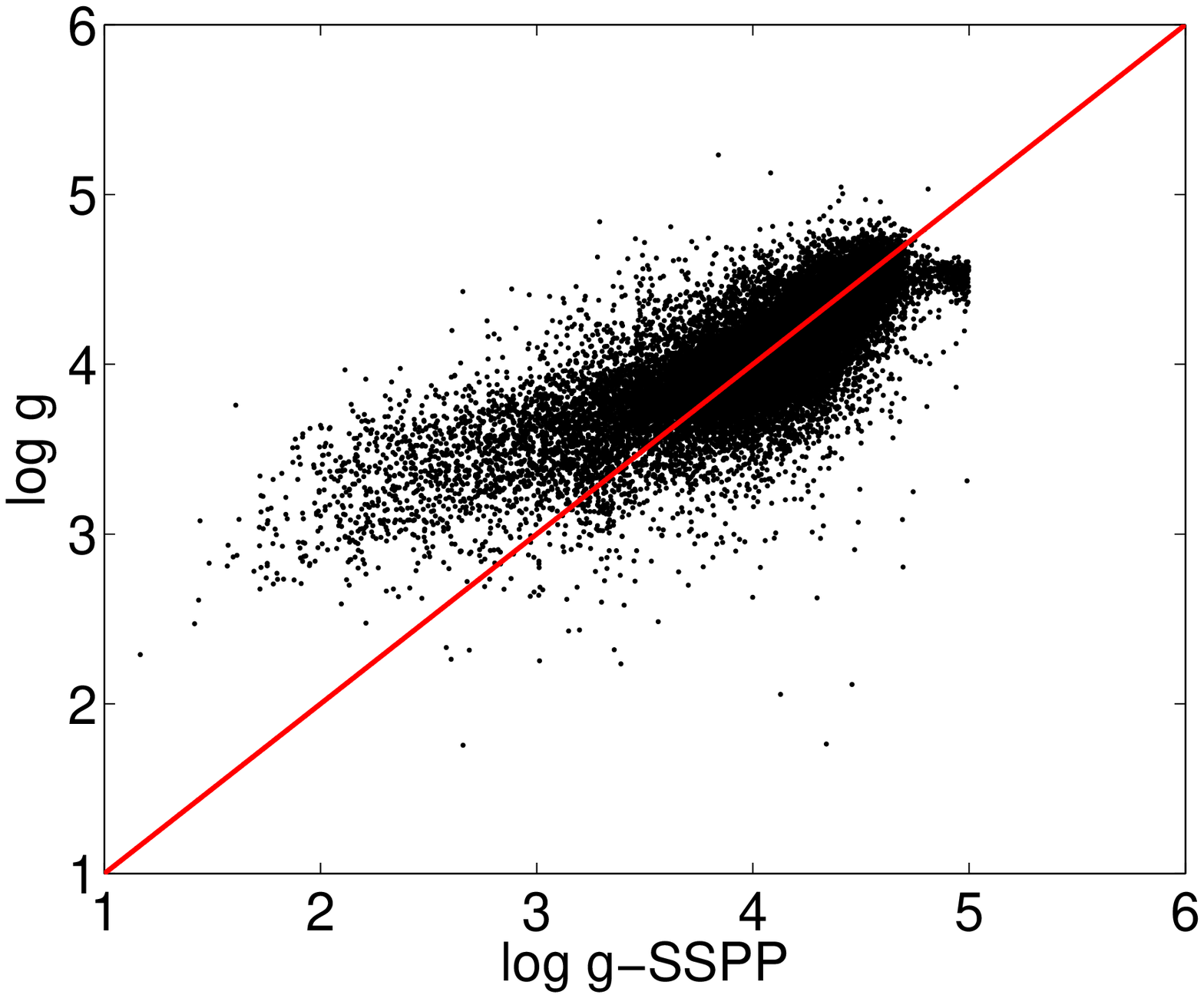} }
  \subfigure[learn from 25000 SDSS spectra for {[Fe/H]}]
    { \includegraphics[height=1.5in,width=2.2in]{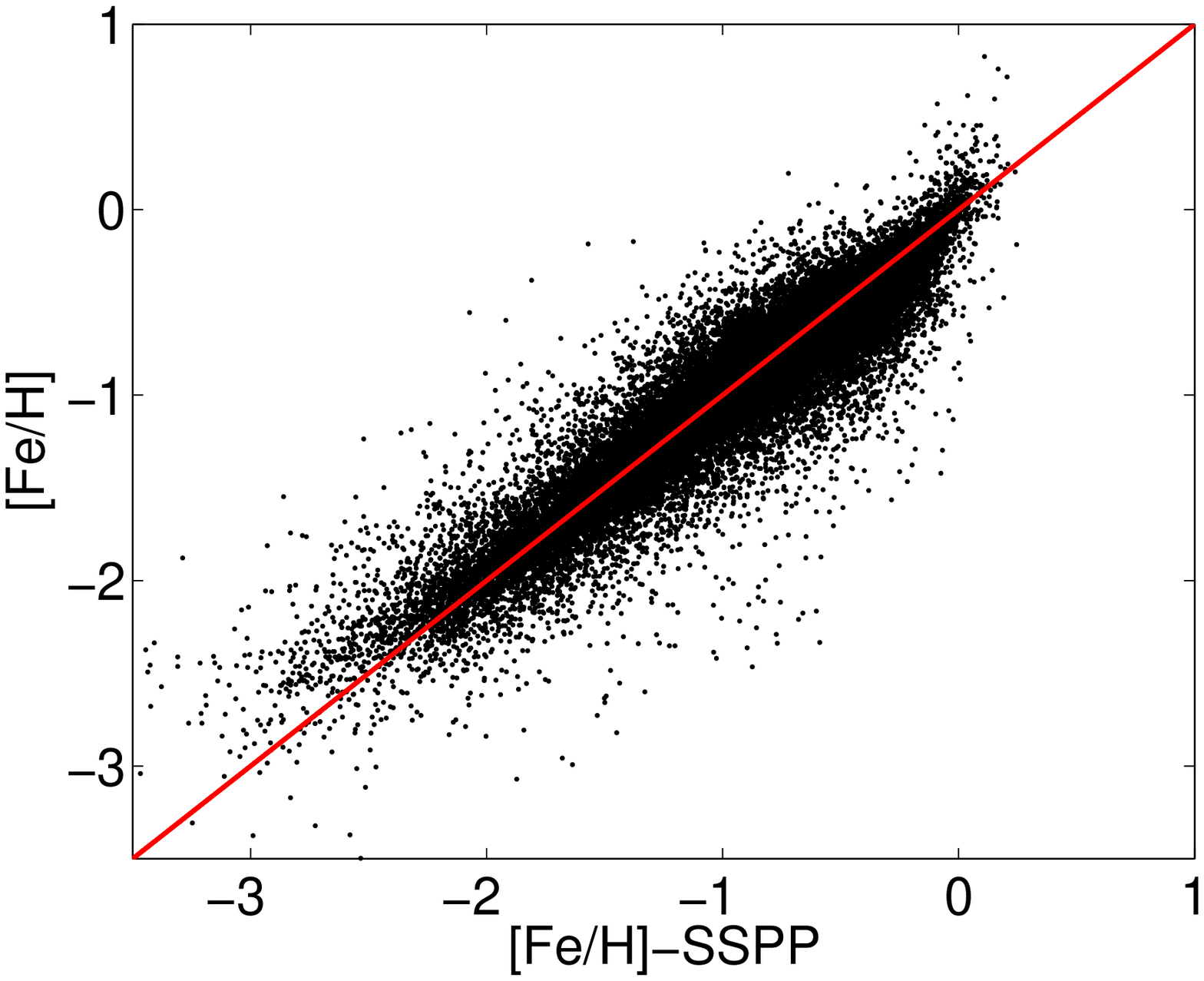} }
  \caption{Dispersion can be improved by increasing the size of training set from 10~000 to 15~000, 20~000, 25~000 in the experiments on SDSS test spectra. This experiment is conducted using SVR$_G$.}
  \label{Fig:comparison:SDSS15000-20000-25000}
\end{figure*}

\begin{figure*}
  \centering
  \subfigure[log T$_\texttt{eff}$  ]
    { \includegraphics[height=1.5in,width=2.2in]{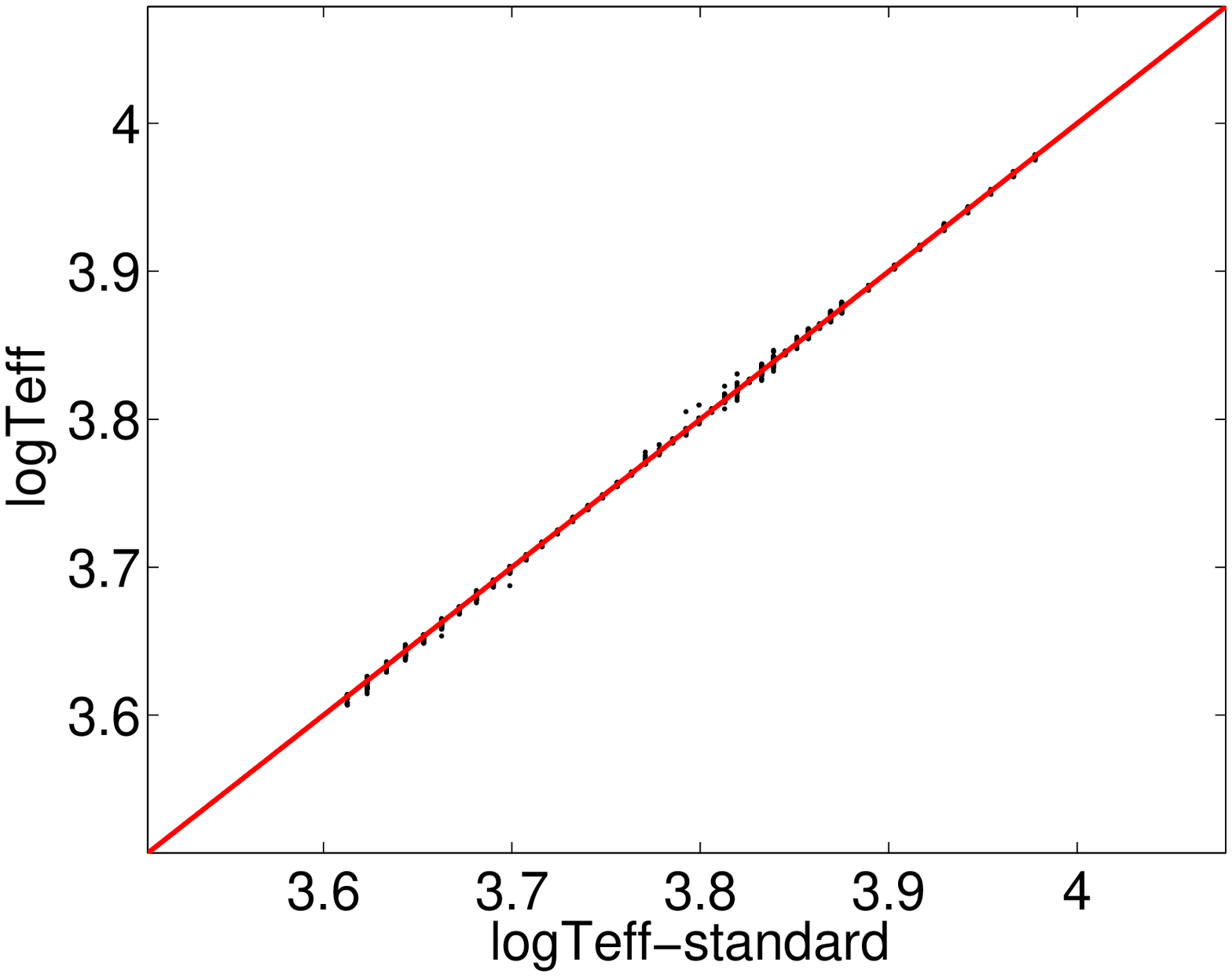} }
  \subfigure[log$~g$]
    { \includegraphics[height=1.5in,width=2.2in]{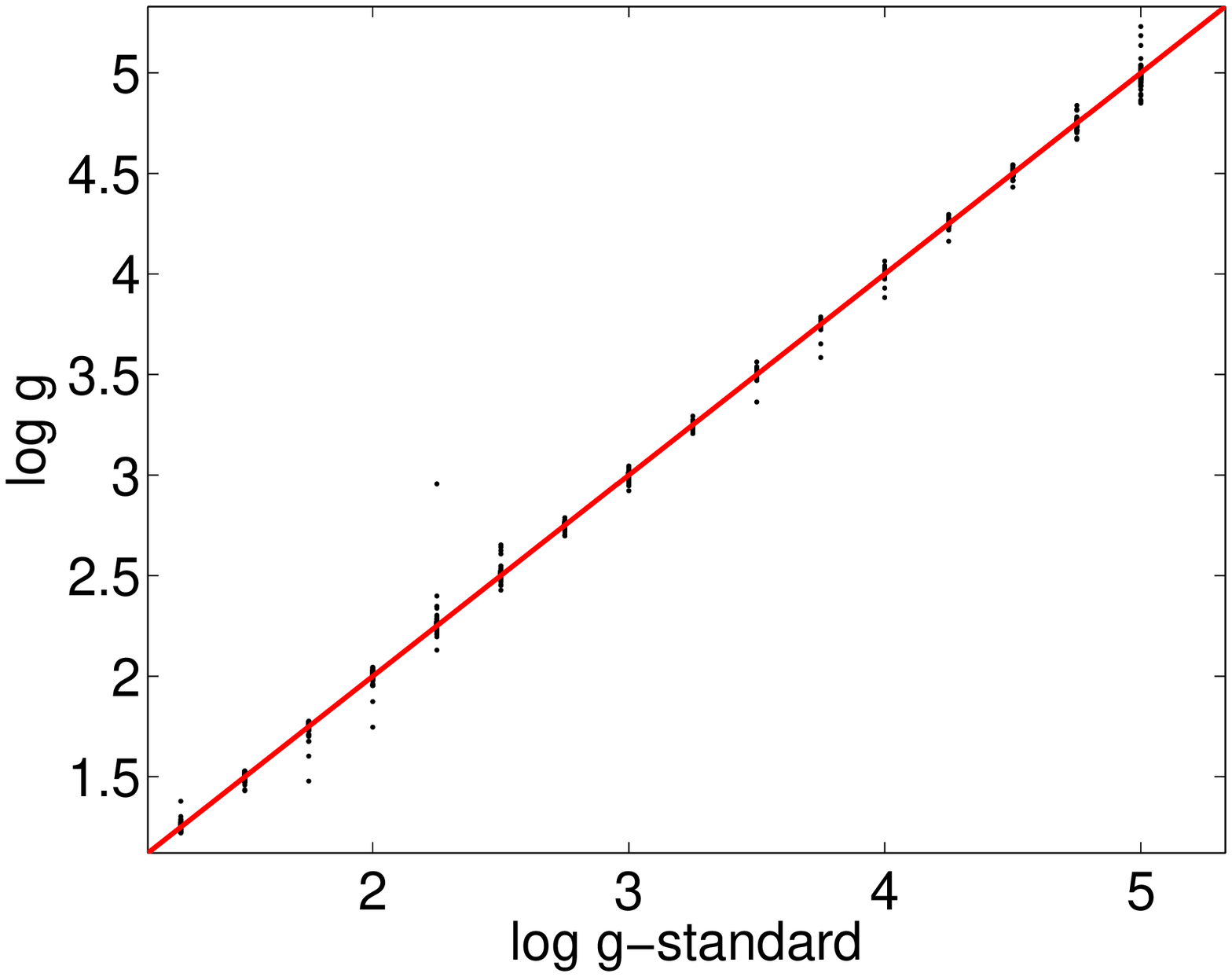} }
  \subfigure[{[Fe/H]}]
    { \includegraphics[height=1.5in,width=2.2in]{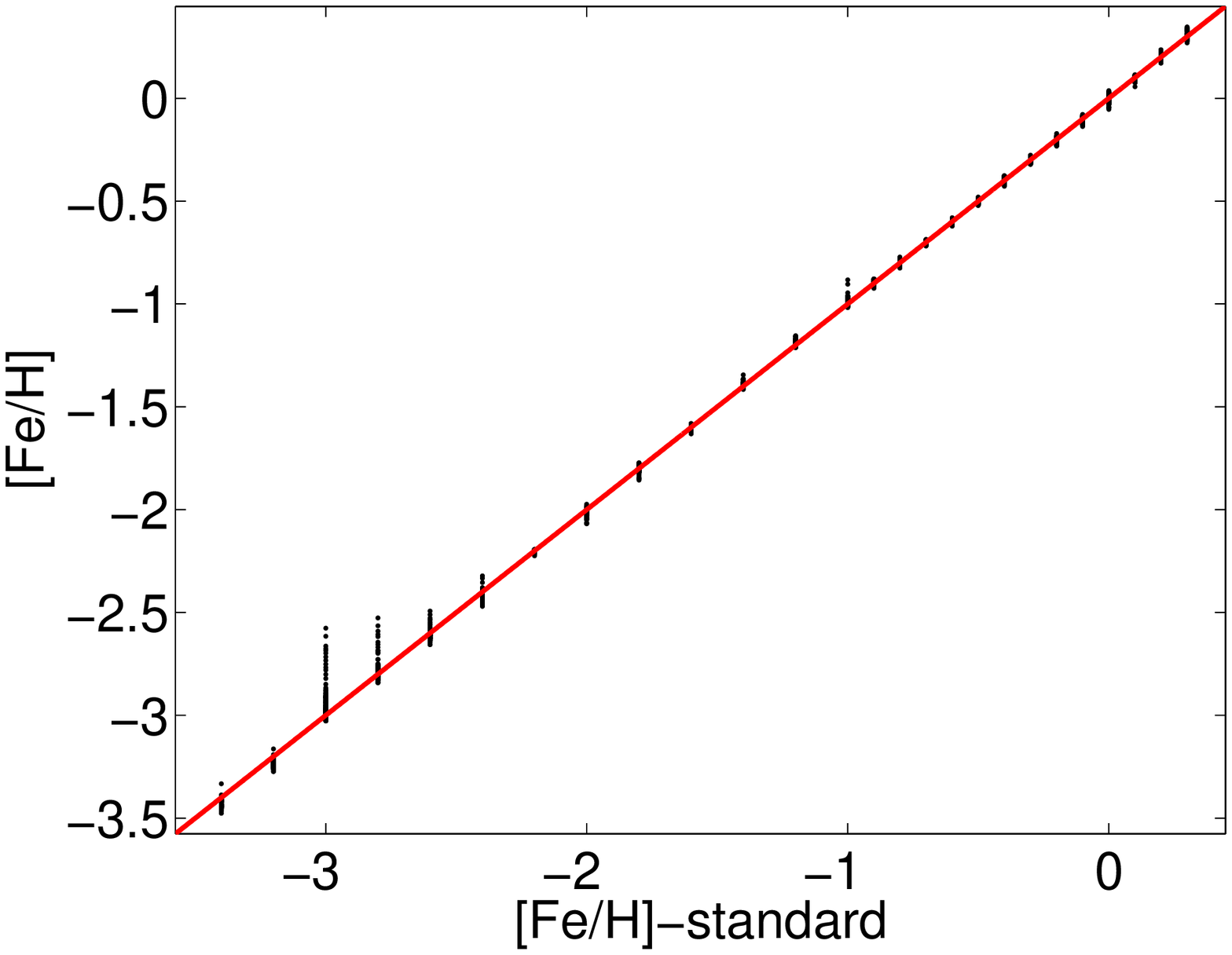} }
  \caption{Dispersion can be improved by increasing the size of training set from 8~500 to 10~469 in the experiment on synthetic test spectra. This experiment is conducted using SVR$_G$.}
  \label{Fig:comparison:KURUCZ}
\end{figure*}

Therefore, the size of the training set affects the performance of the proposed scheme.
For example, if the size of the training set is increased from 10~000 to 15~000, 20~000, 25~000 in the experiments on SDSS spectra\footnote{Correspondingly, the size of the test set decreases from 40~000 to 35~000, 30~000, and 25~000, respectively.}, the test dispersion can be clearly improved (Fig. \ref{Fig:comparison:SDSS15000-20000-25000}). Similar experiments are conducted on synthetic spectra and the corresponding results are presented in Fig. \ref{Fig:comparison:KURUCZ}.

Actuall data usually present some disturbances arising from noise and pre-processing imperfections (e.g., sky lines and/or cosmic ray removal residuals, residual calibration defects and interstellar extinction instability \footnote{By instability, we mean that a slight difference in the interstellar extinction of multiple stars may be observed.}). The negative effect from these factors can be reduced to a certain extent by enriching the knowledge carrier, i.e. the training set (Figs. \ref{Fig:comparison:SDSS15000-20000-25000} and \ref{Fig:comparison:KURUCZ}).

\subsection{Compactness}
For simplicity, this section considers the features of log$~g$ as an example to discuss compactness. Other features of T$_\texttt{eff}$ and [Fe/H] can be discussed similarly.

The original SDSS spectra are described by 3,821 fluxes. To estimate log$~g$, 24 features are detected and data reduction is (3821 - 24)/(3821)$\approx$99.37 percent. This result indicates that log$~g$ can be estimated from a spectral description with a dimension of 24 instead of 3821. The small number of features also implies high efficiency in estimating the parameter from spectral information.

Compactness indicates the study of whether the number of features can be reduced further. Experimental results show that if seven features, namely L1, L8, L13, L19, L20, L21 and L23, are rejected, the feature number decreases by 7/24 = 29 percent and the MAE error increases by 0.0027 dex (approximately 0.0027/0.2035 $\approx$ 1.32 percent).\footnote{These experiments are conducted on SDSS spectra (Section \ref{Sec:Data:SDSS}).} Therefore, the feature number can be refined further if a slight decrease in accuracy is accepted.

\section{Conclusion and Future work}\label{Sec:Conclusion}

This work investigated estimation of effective temperature (T$_\texttt{eff}$), surface gravity (log$~g$) and metallicity ([Fe/H]) from stellar spectra based on the Haar wavelet transform and LASSO algorithm. The proposed scheme is evaluated using actual spectra from SDSS and LAMOST as well as synthetic spectra computed from Kurucz's model. Favorable results are achieved in all cases.

The proposed scheme exhibits excellent robustness and sparseness. The features are extracted using two steps. From the SDSS data, the original spectra, described by 3821 fluxes, are initially decomposed using the Haar wavelet transform and low-frequency coefficients (239 features) are selected as candidate features. In this step, some noises and redundancies with high-frequency components are removed. Then a small subset of the candidate features is chosen as spectral features using the LASSO algorithm. The second step is a supervised learning process that selects features according to their correlation with the parameter to be estimated. The number of selected features is 17 for T$_\texttt{eff}$, 24 for log$~g$ and 25 for [Fe/H]. A representative work is \cite{Journal:Fiorentin:2007}, in which 50 features are extracted for estimating atmospheric parameters.

Another advantage of the proposed scheme is its high accuracy. Using the SVR$_G$ method and 40~000 stellar spectra from SDSS, the MAEs are 0.0062 dex for log~T$_\texttt{eff}$ (85.83~K for T$_\texttt{eff}$), 0.2035 dex for log$~g$ and 0.1512 dex for [Fe/H]. Further details are shown in Table \ref{table:errors:SDSS}. In previous reports, \citet{Journal:Fiorentin:2007} estimated the parameters of 19~000 spectra from SDSS with MAEs of 0.0126 dex for log T$_\texttt{eff}$, 0.3644 dex for log$~g$ and 0.1949 dex for [Fe/H]. \citet{Journal:Jofre:2010} first highly compressed the data using a likelihood method and then estimated the parameters T$_\texttt{eff}$, log$~g$ and [Fe/H] from low-resolution stellar spectra measured by SEGUE; the standard deviations (SDs) of the errors were 130 K for T$_\texttt{eff}$, 0.5 dex for log$~g$ and 0.25 dex for [Fe/H]. Therefore, the results estimated using the proposed scheme exhibit higher accuracy compared with those reported in literature.

In this work, the proposed scheme is evaluated using three different data sets (SDSS, LAMOST and synthetic spectra). On each kind of data, the proposed model is learned and tested independently. An interesting problem is the estimation of atmospheric parameters of one data (e.g., LAMOST ) using a model learned from the other data sets(e.g., SDSS or synthetic spectra). For example, \citet{Journal:Fiorentin:2007} investigated how to estimate the atmospheric parameters of SDSS data from synthetic spectra and vice versa. In this process, residual calibration defects should be considered. This work focuses on sparse feature extraction and the abovementioned problem will be investigated in future work.

\section*{Acknowledgments}
The authors would like to thank the reviewer and editor for their instructive comments and extend their thanks to
Professor Ali Luo and Fang Zuo for their support and discussions.
This work is supported by the National Natural Science Foundation of China (grant No: 61273248, 61075033, 61202315), the Natural Science Foundation of Guangdong Province (2014A030313425,S2011010003348), the Open Project Program of the National Laboratory of Pattern Recognition(NLPR) (201001060) and the high-performance computing platform of South China Normal University..

\appendix

\bsp

\label{lastpage}
%\end{CJK}
\end{document}